\documentclass[11pt]{llncs}
\pdfoutput=1

\usepackage{graphicx}
\usepackage{times}
\usepackage{mathptm}
\usepackage{cite}

\DeclareSymbolFont{AMSb}{U}{msb}{m}{n}
\DeclareSymbolFontAlphabet{\Bbb}{AMSb}
\def\R{\ensuremath{\Bbb R}}
\def\Z{\ensuremath{\Bbb Z}}

\makeatletter
\def\hb@xt@{\hbox to }
\makeatother

\let\oldendproof\endproof
\def\endproof{\qed\oldendproof}

\begin{document}
\title{The Topology of Bendless Three-Dimensional Orthogonal Graph Drawing} 

\author{David Eppstein}

\institute{Computer Science Department\\
University of California, Irvine\\
\email{eppstein@uci.edu}}

\maketitle   

\begin{abstract}
We consider embeddings of 3-regular graphs into 3-dimensional Cartesian coordinates, in such a way that two vertices are adjacent if and only if two of their three coordinates are equal (that is, if they lie on an axis-parallel line) and such that no three points lie on the same axis-parallel line; we call a graph with such an embedding an \emph{$xyz$ graph}. We describe a correspondence between $xyz$ graphs and face-colored embeddings of the graph onto two-dimensional manifolds, and we relate bipartiteness of the $xyz$ graph to orientability of the underlying topological surface. Using this correspondence, we show that planar graphs are $xyz$ graphs if and only if they are bipartite, cubic, and three-connected, and that it is NP-complete to determine whether an arbitrary graph is an $xyz$ graph. We also describe an algorithm with running time $O(n2^{n/2})$ for testing whether a given graph is an $xyz$ graph.
\end{abstract}

\section{Introduction}

\begin{figure}[h]
\centering\includegraphics[width=5in]{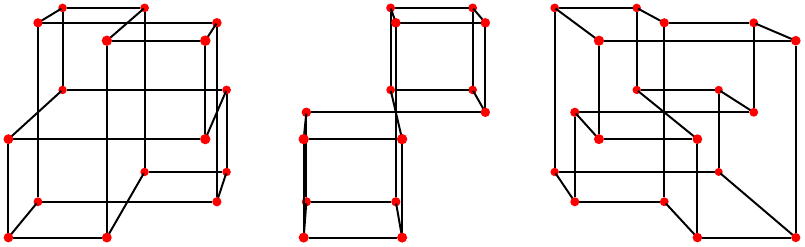}
\caption{The three possible $xyz$ graphs (up to isomorphism) other than the cube with coordinates in $\{0,1,2\}^3$.}
\label{fig:xyz333}
\end{figure}

Consider a finite point set $V$ in $\R^3$ with the following property: every axis-parallel line in $\R^3$ contains either zero or two points of $V$. For instance, the vertices of an axis-aligned cube have this property. Then $V$ defines the vertex set of a cubic (that is, 3-regular) graph, in which two vertices are adjacent if and only if two of their three coordinates are equal; each vertex $v$ is connected to the three other points of $V$ that lie on the three axis-parallel lines through $v$. We call such a graph an \emph{$xyz$ graph}. Figure~\ref{fig:xyz333} depicts the three possible $xyz$ graphs (up to graph isomorphism) other than the cube with coordinates in $\{0,1,2\}^3$.

We are interested in drawings of this type for several reasons.
\begin{itemize}
\item
From the point of view of graph drawing, three-dimensional orthogonal drawings~\cite{EadStiWhi-IPL-96,Woo-GD-98,BieThiWoo-Algo-06,EadSymWhi-GD-96,Woo-01,BieSheWhi-JGAA-99,Woo-TCS-03,PapTol-JGAA-99,CloGarJohWis-JGAA-01} and more generally drawings with very few edge orientations~\cite{DujEppSud-CGTA-07} are significant objects of study. However, past work on three-dimensional orthogonal drawing requires \emph{bends}, in which edges follow axis-aligned polygonal paths rather than simple line segments. An $xyz$ graph provides a particularly simple and well-characterized form of bendless three-dimensional orthogonal drawing. In $xyz$ graphs, unlike much other work on three-dimensional drawing, edges may cross, but this raises no ambiguities in interpreting the drawing, because there are no bends allowed on edges; edge crossings can be distinguished visually from vertices by whether the line segments of the drawing terminate at that point.

\item Placing processors at the points of an $xyz$ graph may be of use in three-dimensional layout of parallel processing intercommunication networks~\cite{CalMas-TCS-01}, providing a layout in which all connected pairs of processors have an open line of sight between each other. For instance, we show that the even-dimensional cube-connected-cycles networks, highly regular cubic graphs of some importance in parallel processing~\cite{PreVui-CACM-81}, have $xyz$ graph layouts.

\item In graph theory, the graphs defined from a planar point set by connecting two points by an edge when they share an $x$- or $y$-coordinate are exactly the line graphs of bipartite graphs, a class of graphs crucial in the characterization of perfect graphs~\cite{ChuRobSey-AM-06}. The $xyz$ graphs are defined from a natural three-dimensional generalization of this construction. As we show, they also have an unexpected connection to topological graph theory~\cite{GroTuc-87,BonLit-95} and graph coloring~\cite{NelWil-90,JenTof-95}, in that any $xyz$ graph corresponds to a three-coloring of the faces of an embedding of a graph on a 2-manifold; such face-colored embeddings arise naturally from the GEM (graph-embedded map) representation of arbitrary 2-manifold embeddings of graphs~\cite{BonLit-95,Epp-SODA-03}. And, since $xyz$ graphs form a naturally-defined class of cubic graphs, greater understanding of $xyz$ graphs has the possibility of shedding light on other cubic graph classification problems, such as that of enumerating the cubic partial cubes~\cite{Epp-EJC-06}.
\end{itemize}

\section{New results}

\begin{itemize}
\item We prove an equivalence between the graphs that can be embedded in $\R^3$ as $xyz$ graphs and the graphs that can be embedded on two-dimensional surfaces as the boundaries of certain 3-face-colored cell complexes, which we call $xyz$ surfaces. Using this equivalence, we may use topological tools to study the existence of $xyz$ graph embeddings; for instance we show that an $xyz$ graph is bipartite if and only if the corresponding $xyz$ surface is orientable.
\item We provide several examples of $xyz$ graphs, including the graph-encoded maps representing surface embeddings of graphs, the cube-connected cycles of even order, and the skeletons of polyhedra with even faces.
\item We show that it is NP-complete to determine whether a given graph can be embedded in $\R^3$ as an $xyz$ graph.
\item We provide an algorithm with running time $O(n2^{n/2})$ for determining whether a given graph can be embedded in $\R^3$ as an $xyz$ graph.
\item We show that any cubic graph can be covered by an $xyz$ graph and that any cubic map can be covered by an $xyz$ surface.
\end{itemize}

\section{Topology of $xyz$ graphs}

If $G$ is an undirected graph, and $C$ is a multiset of simple cycles in $G$, we may define a \emph{cell complex} as a disjoint union of points, line segments, and disks: one point for each vertex, one line segment for each edge, and one disk for each cycle, glued together topologically according to the connection pattern given by $G$. For instance, if $G$ is a cube graph (having eight vertices and twelve edges), and $C$ is the set of four-cycles in $G$, the surface we get can be realized as the set of vertices, edges, and facets of a geometric cube. However, complexes of this type may be defined independently of any embedding of the whole complex into three-dimensional space. If, further, every point on the cell complex has a neighborhood that is topologically equivalent to an open disk, it is called a \emph{2-manifold} (without boundary). In graph theoretic terms, the cell complex is a manifold whenever two conditions are satisfied:
\begin{enumerate}
\item Each edge of $G$ must belong to exactly two cycles of $C$.
\item At each vertex $v$ of $G$, one can reach any incident edge from any other incident edge by a chain of edge-face-edge steps in which each edge and face is also incident to $v$.
\end{enumerate}
The second condition prevents nonmanifold complexes such as those formed by two polyhedra that meet at a single vertex. If $G$ is a cubic graph, the second condition is satisfied automatically, and only the first condition is needed. We call the cell complex defined from $G$ and $C$ an \emph{embedding} of $G$ onto a manifold, and we call the cycles of $C$ the \emph{faces} of the embedding.
\emph{Topological graph theory}~\cite{GroTuc-87,BonLit-95}  studies embeddings of graphs onto manifolds of this type.

We define an \emph{$xyz$ surface} to be an embedding of a cubic graph $G$ onto a manifold, defined by a collection of faces $C$, with the following additional properties.
\begin{enumerate}
\item Any two faces intersect in either a single edge of $G$ or the empty set.
\item The faces of $C$ can be assigned three colors such that no two faces sharing an edge have the same color.
\end{enumerate}

The first of these two properties is commonly referred to in the topological graph theory literature by saying that the embedding must be \emph{polyhedral} (see., e.g., \cite{Koc-GD-08}). For non-cubic graphs, polyhedral embeddings may also include pairs of faces that intersect in a single vertex, but this cannot happen in a cubic graph. Craft and White~\cite{CraWhi-DM-08} have studied a similar 3-coloring condition on orientable maps without the polyhedral condition.

\begin{figure}[t]
\centering\includegraphics[width=5in]{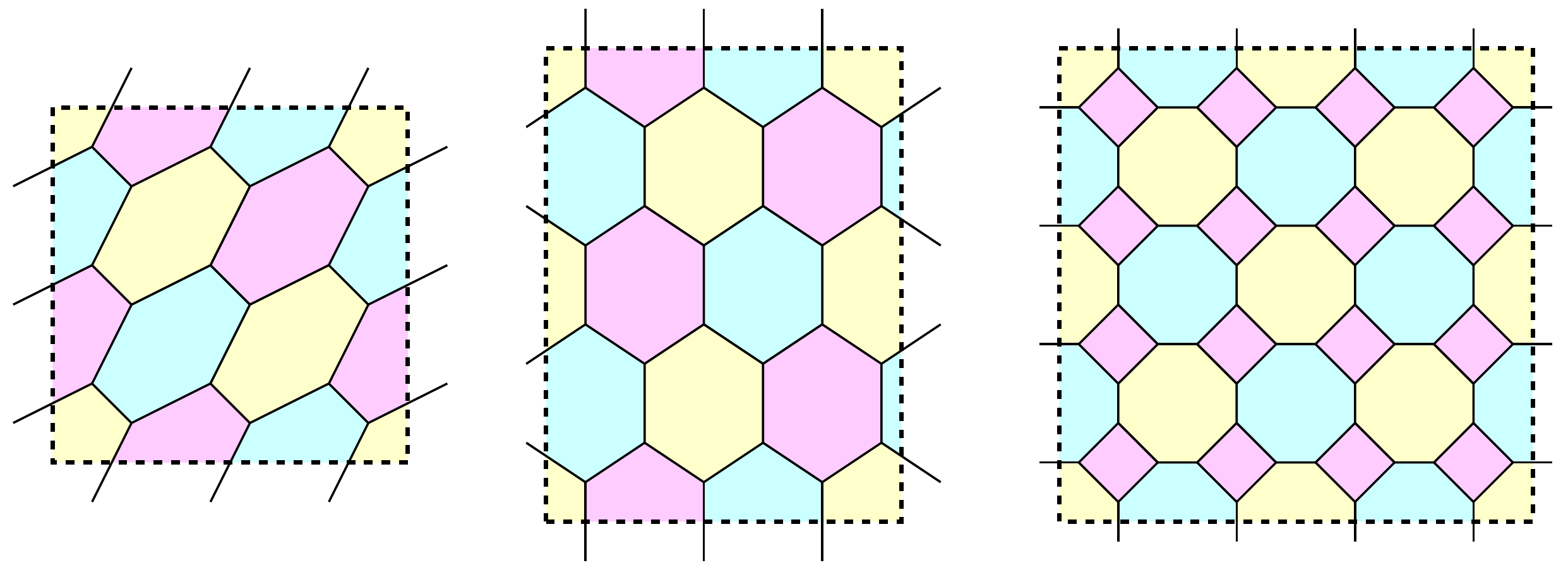}
\caption{Three $xyz$ surfaces, each with the topology of the torus. In each case, the torus is depicted as cut and unrolled into a rectangle; the corresponding topological surface is formed by gluing opposite pairs of rectangle edges.}
\label{fig:3x-tori}
\end{figure}

Figure~\ref{fig:3x-tori} depicts three $xyz$ surfaces, all topologically equivalent to tori. The leftmost is an embedding of the Pappus graph onto a torus, with nine faces and 18 vertices; the same graph is shown as the rightmost graph of Figure~\ref{fig:xyz333}. The middle surface shows the embedding of a graph with twelve faces, 24 vertices, and 36 edges. The right surface is a torus embedding of the 64-vertex four-dimensional cube-connected cycles network.

\begin{theorem}
\label{thm:xyz surface}
A cubic graph $G$ is an $xyz$ graph if and only if $G$ can be embedded as an $xyz$ surface.
\end{theorem}

\begin{proof}
Suppose that $G$ is an $xyz$ graph; we must show that $G$ can be embedded as an $xyz$ surface.
Fix a particular $xyz$ graph representation of $G$.
We let $C$ consist of the cycles in $G$ that use only two of the three orientations in the $xyz$ graph representation; that is, each such cycle lies in a plane parallel to two coordinate axes. Each edge of $G$ belongs to two such cycles, one for each of the coordinate planes to which it is parallel; therefore, since $G$ is cubic, $C$ forms an embedding of $G$ onto a manifold. The cycles of $C$ can be colored according to the coordinate planes they are parallel to. Since the cycles of $G$ in any single coordinate plane are disjoint, two cycles can have a nonempty intersection only if they belong to different planes; in that case, the intersection must lie on the axis-parallel line formed by the intersection of the two planes containing the cycles, and consists of the edge of $G$ that lies on that same line.

Conversely, suppose that $G$ is embedded as an $xyz$ surface, with cycle set $C$; we must find an $xyz$ graph representation for $G$. Let $X$, $Y$, and $Z$ be the three monochromatic subsets of $C$, and let the faces in $C$ be numbered $f_0, f_1, \dots$. Each vertex $v$ in $G$ is incident to exactly three faces: $f_i$ in $X$, $f_j$ in $Y$, and $f_k$ in $Z$ for some $i,j,k$. We assign $v$ the three-dimensional coordinates $(i,j,k)$. If two vertices $u$ and $v$ are adjacent, they share the two coordinates determined by the two faces containing edge $uv$, and therefore lie on an axis-parallel line of the embedding of $G$ into $\R^3$. Conversely, if two vertices $u$ and $v$ are not adjacent, they can lie on at most one face of $C$ (else the faces they lie on would intersect in more than one vertex) and therefore have at most one coordinate in common, so they do not lie on an axis-parallel line. Thus, the three axis-parallel lines through each embedded vertex $v$ each contain exactly two points, which are the neighbors of $v$, so the embedding forms an $xyz$ graph representation of $G$.
\end{proof}

\begin{figure}[t]
\centering\includegraphics[height=1.25in]{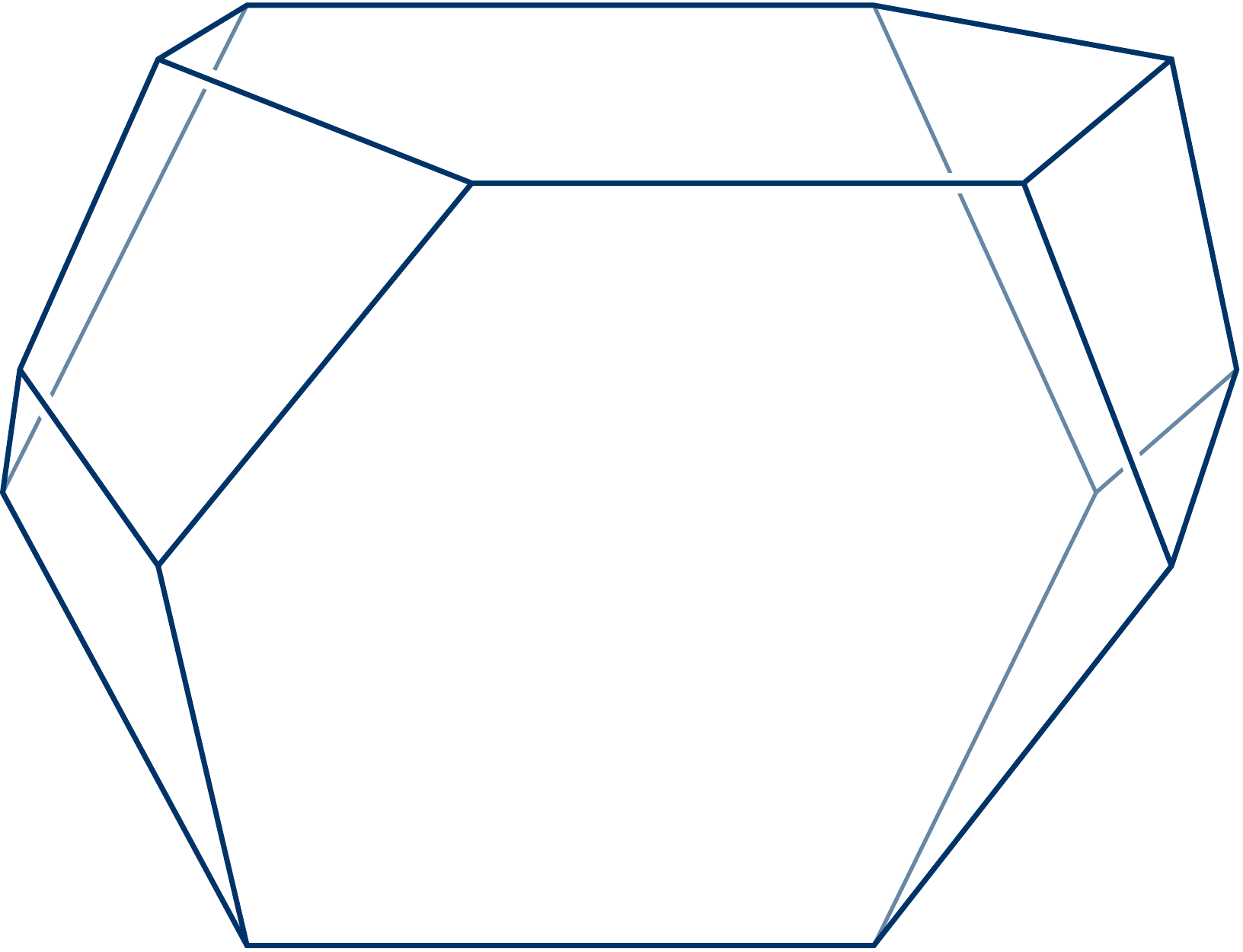}
\caption{Nine-sided polyhedron with the same face structure as the $xyz$ surface formed from the graph in the center of Figure~\ref{fig:xyz333}.}
\label{fig:nonahedron}
\end{figure}

For instance, we can conclude from this result and the rightmost $xyz$ surface in Figure~\ref{fig:3x-tori} that the four-dimensional cube-connected cycle graph is an $xyz$ graph. In the other direction, the three $xyz$ graphs depicted in Figure~\ref{fig:xyz333} form $xyz$ surfaces that are (in left to right order) a surface on the projective plane (resembling the Roman surface in its geometric arrangement), a spherical surface combinatorially equivalent to a polyhedron with three hexagonal facets and six quadrilateral facets (Figure~\ref{fig:nonahedron}), and an embedding of the Pappus graph on the torus combinatorially equivalent to the leftmost $xyz$ surface of Figure~\ref{fig:3x-tori}.

The embedding of Theorem~\ref{thm:xyz surface} can be used to embed any $xyz$ graph into an $\frac{n}4\times\frac{n}4\times\frac{n}4$ grid. To see this, observe that each face of an $xyz$ graph or an $xyz$ surface must have even length, because it alternates between edges parallel to two different coordinate axes in the case of an $xyz$ graph, and between edges incident to pairs of faces with two different sets of colors in the case of an $xyz$ surface. Since the minimum length of a face in an embedding of a simple graph, three, is not an even number, all faces must have length at least four. Thus, the number of faces in
in any color class of an $xyz$ surface coloring is at most $n/4$: each vertex belongs to one face of that color, but each face contains at least four vertices. Since each face provides a value for one of the coordinates in the grid embedding, the number of distinct values for each coordinate is therefore also at most $n/4$.
However this bound is tight only for the cube, the only $xyz$ surface in which all faces are quadrilaterals: any other $xyz$ surface has a face with more than four vertices, and therefore has a color class with fewer than $n/4$ faces, leading to an embedding on a grid with fewer than $n/4$ distinct values in one of the coordinates. For many graphs, it is possible to find multiple $xyz$ graph embeddings that differ geometrically, although they are combinatorially and topologically equivalent, by permuting the coordinate values that correspond to the faces of $C$. Additionally, in some cases, smaller grids may be obtained by using equal coordinate values for multiple faces of the same color. We do not consider problems of choosing these coordinate values in order to improve some aspect of the resulting graph drawing in this paper, but such problems are a natural subject for future work.

For many of the subsequent results, we will find it more convenient to work topologically, in terms of $xyz$ surfaces, and less convenient to work geometrically, in terms of an explicit $xyz$ graph representation. However, in situations where an $xyz$ graph representation is desired (e.g. as the output of a graph drawing algorithm) it may easily be reconstructed using the face-numbering technique of the proof above.

\section{Properties of $xyz$ graphs}

\begin{theorem}
Every $xyz$ graph is triangle-free.
\end{theorem}

\begin{proof}
The edges of a triangle must be mutually perpendicular, as each pair of edges meets at a vertex. But then, if one follows a path around the triangle starting from one of its vertices, in an $xyz$ graph embedding, each step of the path changes one of the three coordinates, and each coordinate is changed exactly once, so the path cannot return to its starting point, a contradiction.
\end{proof}

\begin{lemma}
\label{lem:3con}
Let $abc$ be a path of three vertices in an $xyz$ graph $G$, and let $d$ be a vertex of $G$ distinct from $a$, $b$, and $c$. Then there exists a path in $G$ that starts at $a$, ends at $c$, and does not pass through either $b$ or $d$.
\end{lemma}

\begin{proof}
If $d$ does not belong to the face of $G$ that contains edges $ab$ and $bc$, we may choose as our path the complementary set of edges in the same face. Otherwise, let $e$ be the third neighbor of $b$ in $G$, and form a path by concatenating the complementary set of edges to edges $ab$ and $be$ in the face shared by those two edges, together with the complementary set of edges to edges $bc$ and $be$ in the face shared by those two edges. Neither of these two faces can contain $d$, as then they would intersect the face containing $ab$ and $bc$ in more than a single edge.
\end{proof}

\begin{theorem}
\label{thm:3con}
Every connected $xyz$ graph is 3-vertex-connected.
\end{theorem}

\begin{proof}
Let $s$ and $t$ be two vertices in an $xyz$ graph $G$ from which two other vertices $u$ and $v$ have been deleted. We must show that there still remains a path from $s$ to $t$. Let $P$ be a path in $G$ that connects $s$ to $t$. If $P$ contains $u$, we may apply Lemma~\ref{lem:3con} to the segment of $P$ formed by the two edges incident to $u$, replacing it by another path that avoids $u$. Similarly, if $P$ contains $v$, we may apply Lemma~\ref{lem:3con} again, replacing it by another path that avoids $v$ without adding any additional component of path through $u$. Thus, since we can connect any two vertices after the removal of any other two vertices, the graph is 3-vertex-connected.
\end{proof}

We conclude this section with an interesting connection between bipartiteness and topology. An \emph{orientation} of a graph embedding onto a surface, described by a set of cycles $C$, can be described as a choice of orientation for each edge in each cycle, such that each cycle of $C$ is given a consistent cyclic orientation, and such that the two cycles shared by any edge $e$ assign opposite orientations to $e$. A surface is \emph{orientable} if graphs embedded on it may be oriented in this way; for instance the sphere and torus are orientable, while the projective plane is not.

\begin{theorem}
\label{thm:orientability}
Let $G$ be a graph embedded onto an $xyz$ surface. Then $G$ is bipartite if and only if the surface is orientable.
\end{theorem}

\begin{proof}
Let $G$ be a bipartite $xyz$ graph; we must show that the corresponding $xyz$ surface is orientable. Color the vertices of $G$ black and white, then we can orient the faces in $xy$-parallel planes from black to white in the $x$ direction and from white to black in the $y$ direction, in $xz$-parallel planes from white to black in the $x$ direction and from black to white in the $z$ direction, and in $yz$-parallel planes from black to white in the $y$ direction and from white to black in the $z$ direction; the result is a consistent orientation of all faces of the graph.

Conversely, suppose that $G$ is embedded as an $xyz$ graph on an orientable surface; we must show that $G$ is bipartite. Once we orient all faces of the graph, the same correspondence described above can be used to color the vertices of $G$; the consistency of the orientation will lead to a consistent two-coloring of the graph.
\end{proof}

A standard result in topology is that 2-manifolds may be classified by their orientability and their \emph{Euler characteristic} $|V|-|E|+|C|$, so from Theorem~\ref{thm:orientability} it is straightforward to determine the topological type of any $xyz$ embedding by computing the Euler characteristic of the embedding and testing the bipartiteness of the graph.

\section{Algorithms for $xyz$ embedding}

As we now show, there exist efficient algorithms to determine whether an embedded surface is an $xyz$ surface, or whether a partition of the edges of a graph into three perfect matchings can be used as the three parallel classes of edges in an $xyz$ graph. However, it is not so easy to find an $xyz$ graph representation for an initially unlabeled graph.

\begin{theorem}
\label{thm:test xyz surface}
Let $G$ be a connected undirected $n$-vertex graph, and let $C$ be a collection of cycles in $G$. Then in time $O(n)$ we may determine whether $C$ is the set of cycles of an $xyz$ surface embedding of $G$, and if so construct an $xyz$ graph representation of $G$.
\end{theorem}

\begin{proof}
We first check that $G$ is a cubic graph and that $C$ covers each edge of $G$ exactly twice.
Next, we assign arbitrary index numbers to the cycles in $C$. Each edge has an associated pair of index numbers, which we order lexicographically. We may sort the edges of $G$ according to this lexicographic ordering in linear time, by two passes of bucket sorting; in the resulting sorted order, if some pairs of edges both belong to the same two cycles, at least one such pair will appear consecutively. Thus, by performing this sorting algorithm and then testing adjacent pairs of edges in the sorted order, we may verify in linear time that each pair of cycles intersects in at most a single edge.

To test 3-colorability of the cycles in $C$, we apply the following algorithm. We store a set of the available colors for each cycle (initially, all three colors are available for each cycle), and a list $L$ of cycles that have only one remaining color. We will color the cycles in some order; whenever we color a cycle we remove that color from the available colors of all cycles that share an edge with it, and update $L$ whenever that removal causes an adjacent cycle to have only one remaining available color. We begin this sequence of color choices by choosing arbitrarily two cycles that share an edge, and assigning arbitrarily two different colors to those two cycles. Then, as long as $L$ remains nonempty, we remove a cycle from $L$, and assign it the one color that is available to it.

If this process terminates with a 3-coloring of all faces in $C$, we have found an $xyz$ surface representation for $G$. Conversely, suppose that $G$ has an $xyz$ surface representation: we argue that this process will necessarily find a correct 3-coloring of all faces. To show this, permute the colors of the coloring if necessary so that they match the colors chosen for these faces at the start of the algorithm. Clearly, every color choice subsequent to that is forced, so the algorithm can never choose an incorrect color for a face, and therefore also can never eliminate the correct color for any face; the only way it could fail to 3-color all faces would be if it terminated with $L$ empty before coloring all faces.
But if $f$ is any face of $C$, let $p$ be any path connecting a vertex of the shared edge of the first two colored faces with any vertex of $f$. At any stage in the algorithm until $f$ has been colored, let $v$ be the vertex of $p$ that is closest along the path to the first two colored faces, and that is incident to an uncolored face $f'$; then the two differently-colored neighboring faces of $f'$ at $v$ would force $f'$ to belong to $L$. Thus, $L$ cannot be empty until $f$ is colored, but since this is true for any face $f$ the algorithm cannot terminate when given as input a 3-colorable surface embedding until all faces are colored.
\end{proof}

\begin{corollary}
\label{cor:test xyz graph}
Let $G$ be a connected undirected $n$-vertex graph, and let $E_1$, $E_2$, $E_3$ be a partition of the edges of $G$ into three matchings. Then in time $O(n)$ we may determine whether there is an $xyz$ graph representation of $G$ in which each set $E_i$ is the set of edges parallel to the $i$th coordinate axis.
\end{corollary}

\begin{proof}
For each pair $E_i$ and $E_j$, $E_i\cup E_j$ is a disjoint union of cycles; we let $C$ be the set of cycles formed in this way for all three pairs of matchings, and apply Theorem~\ref{thm:test xyz surface} to this set of cycles.
\end{proof}

\begin{lemma}
\label{lem:all-match-partitions}
Let $G$ be a biconnected cubic graph. Then there are at most $2^{(n-2)/2}$ partitions of the edges of $G$ into three perfect matchings, and these partitions may be listed in time $O(2^{n/2})$.
\end{lemma}

\begin{proof}
We compute an $st$-numbering of $G$~\cite{EveTar-TCS-76}; that is, an ordering of the vertices of $G$ in which each vertex, except for the ones at the start and the end of the sequence, has a neighbor that occurs earlier in the sequence and a neighbor that occurs later in the sequence. We define a \emph{split vertex} to be one with one previous neighbor and two later neighbors, and a \emph{merge vertex} to be one with two previous neighbors and one later neighbor. If there are $k$ split vertices there would be $3+2k+(n-k-2)$ edges, as the first vertex in the $st$-numbering is the earlier endpoint of three edges, the split vertices are each the earlier endpoint of two edges, the $n-k-2$ merge vertices are each the earlier endpoint of only one edge, and the final vertex in the $st$-numbering is the earlier endpoint of no edges. Observing that the graph has $3n/2$ edges total and solving for $k$, we find that there must be exactly $(n-2)/2$ split vertices.

To list all partitions, we then perform a backtracking algorithm in which we assign the edges to partitions in order by their earlier endpoints in the $st$-numbering; once we make an assignment for an edge $e$ we recursively list all partitions for edges occurring later in this ordering before backtracking and trying an alternative assignment for $e$ (if an alternative exists). If this backtracking process ever reaches a contradictory state in which no possible assignment is available from an edge, it backtracks without recursing.

At the initial vertex of the $st$-numbering, the backtracking algorithm has no choices to make: it can partition the incident edges into three disjoint subsets in only one way. At the final vertex, there is again no choice to make, because all incident edges must already have been partitioned. And at each merge vertex, there is no choice to make, because there are two incident edges which must already have been placed into two sets of the partition, and the third incident edge can only go in the third set of the partition. Thus, the only branch points of this backtracking algorithm are the split vertices, at which the two edges for which the vertex is the earlier endpoint must be assigned to the two remaining partition sets, in either of two different ways.

Since the algorithm makes a binary choice at each of $(n-2)/2$ levels of its recursion, its total time is $O(2^{n/2})$. The number of partitions listed is at most the number of leaves in a binary tree of height $(n-2)/2$, which is $2^{(n-2)/2}$.
\end{proof}

Greg Kuperberg (personal communication) has pointed out that the prisms over $n/2$-gons form biconnected cubic graphs with $\Omega(2^{n/2})$ partitions into three perfect matchings, showing that this bound is tight to within a constant factor.

\begin{theorem}
We can test whether a given unlabeled graph is an $xyz$ graph, and if so find an $xyz$ graph representation of it, in time $O(n2^{n/2})$.
\end{theorem}

\begin{proof}
We list all partitions into matchings using Lemma~\ref{lem:all-match-partitions}, and test whether any of them can be used to define an $xyz$ graph representation using Corollary~\ref{cor:test xyz graph}.
\end{proof}

We have implemented our algorithms for listing all partitions of a cubic graph into perfect matchings and for testing whether a given graph is an $xyz$ graph, using the Python programming language. The implementation is available online at http://www.ics.uci.edu/$\sim$eppstein/PADS/xyzGraph.py.

\section{Cayley graphs}

\begin{figure}[t]
\centering
\includegraphics[height=3in]{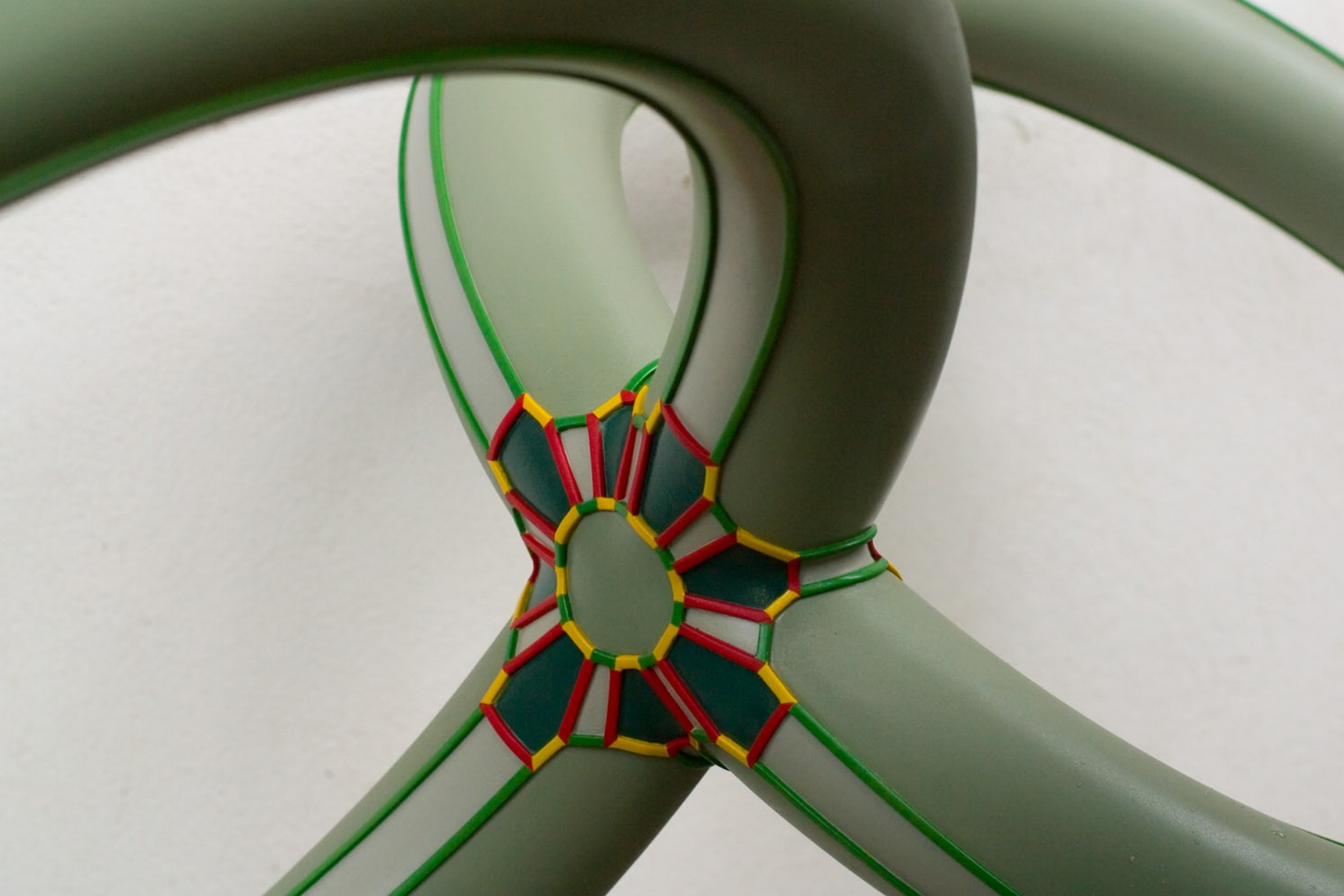}
\caption{The Cayley graph of the automorphism group of the M\"obius--Kantor graph $G(8,3)$, embedded on a double torus~\cite{Tuc-JCTB-84}.  The partition of the faces into red-green rectangles, red-yellow hexagons, and yellow-green 16-gons shows that this embedding is an $xyz$ surface. Sculpture ``Tucker's Genus Two Group'' by DeWitt Godfrey and Duane Martinez, at the Technical Museum of Slovenia; photograph by the author.}
\label{fig:TuckerGroup}
\end{figure}

\begin{figure}[t]
\centering\includegraphics[width=2.5in]{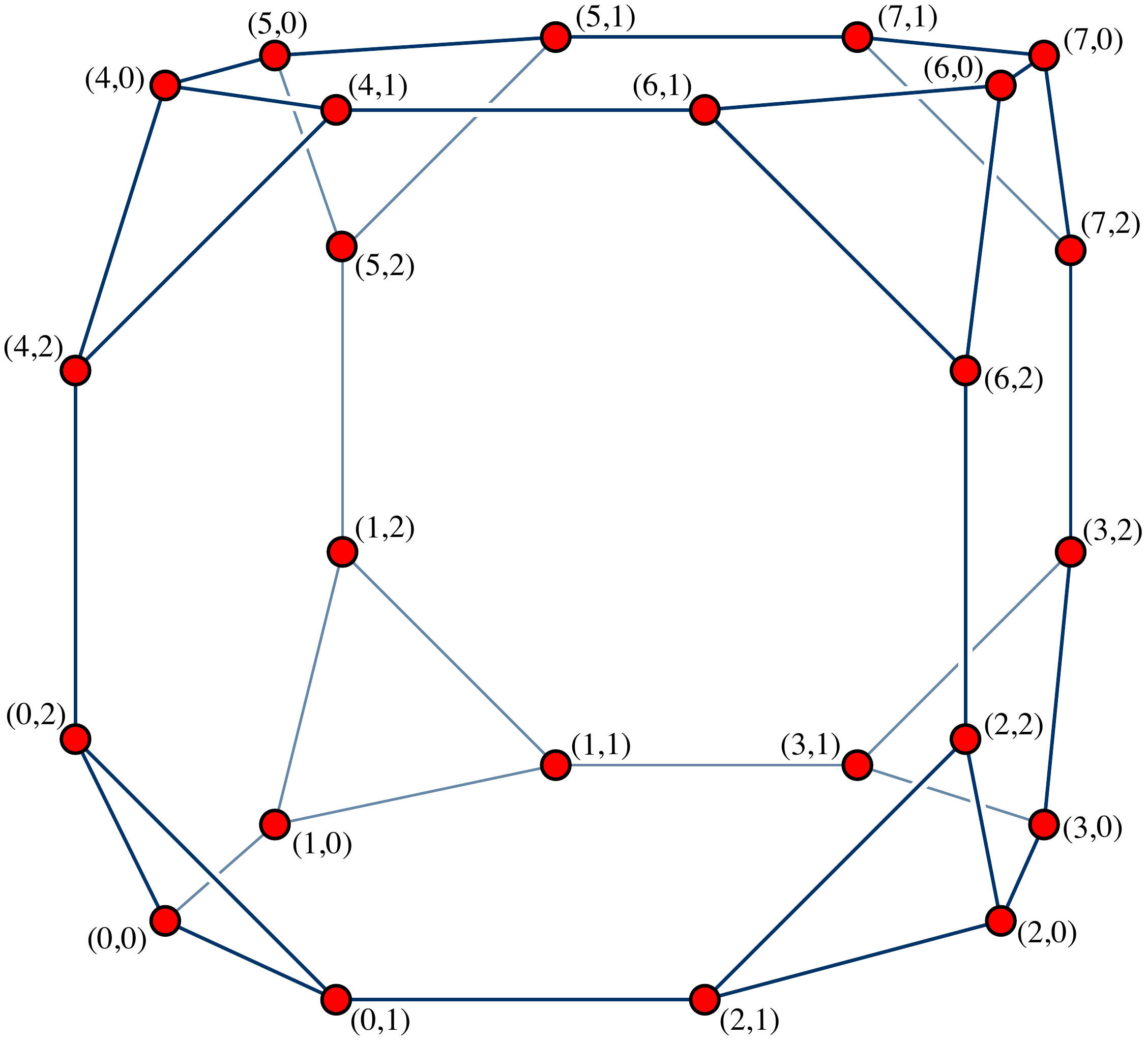}
\caption{The cube-connected cycles of order 3, not an $xyz$ graph because it contains triangles.}
\label{fig:ccc3}
\end{figure}

A \emph{Cayley graph} is a graph having as its vertices the members of a finite group, and its edges determined by a subset of \emph{generators} for that group; there is an edge from $g$ to $gs$ whenever $g$ is a group element and $s$ is one of the chosen generators.  An example is shown in Figure~\ref{fig:TuckerGroup}, the Cayley graph of a 96-element group embedded on a double torus. This group was shown by Tucker~\cite{Tuc-JCTB-84} to be the unique group for which the minimal genus of any embedding of a Cayley graph on a surface is two. The colors of the edges in the figure show the partition of the edges of the graph according to which generator they are formed by; each generator is self-inverse. The embedding shown in the figure is an $xyz$ surface: the faces of the embedding are bounded by edges formed by only two of the three generators, and can be partitioned into three color classes according to which two generators form the edges bounding each face. Therefore, this Cayley graph is an $xyz$ graph.

Another example, the cube-connected cycles network $CCC_n$, of importance in parallel processing~\cite{PreVui-CACM-81}, is formed by replacing every vertex of a hypercube by a cycle. It consists of $n2^n$ nodes that can be indexed by pairs of numbers $(x,y)$ where $0\le x<2^n$ and $0\le y<n$; the neighbors of node $(x,y)$ are $(x,(y+1)\bmod n)$, $(x,(y-1)\bmod n)$, and $(x\oplus 2^y,y)$, where $\oplus$ denotes bitwise exclusive or; this is a Cayley graph for the group of operations on $n$-bit binary words generated by single-bit rotations of the word and flips of the first bit of the word~\cite{AnnBauRos-90}. The cube-connected cycles of order three (Figure~\ref{fig:ccc3}) cannot be an $xyz$ graph, as it is not triangle-free, but we have already seen (Figure~\ref{fig:3x-tori}, right) that the cube-connected cycles of order four is an $xyz$ graph.

\begin{theorem}
Let $n$ be any even number greater than or equal to four. Then the cube-connected cycles network $CCC_n$ is an $xyz$ graph.
\end{theorem}

\begin{proof}
We define a set of faces $C$, consisting of the $n$-vertex cycles $c_x=\{(x,y)\mid 0\le y<n\}$ for each $x$ in the range $0\le x<2^i$, and the eight-vertex cycles $e_{x,i}$ consisting of the eight vertices $(x,i)$, $(x\oplus 2^i, i)$, $(x\oplus 2^i,j)$, $(x\oplus 2^i\oplus 2^j,j)$, $(x\oplus 2^i\oplus 2^j,i)$, $(x\oplus 2^j,i)$, $(x\oplus 2^j, j)$, and $(x,j)$,
for each $x$ in the range $0\le x<2^i$, each $i$ in the range $0\le i<n$, and $j = i + 1$ (mod $n$).
Note that $e_{x,i} = e_{x\oplus 2^i, i} = e_{x\oplus 2^i\oplus 2^j,i} = e_{x\oplus 2^j,i}$, and that each edge of $CCC_n$ belongs to exactly two cycles among the cycles $c_x$ and $e_{x,i}$.
This collection of faces defines an embedding of any cube-connected cycles onto a 2-manifold, regardless of whether $n$ is odd; for instance the truncated cube shown in Figure~\ref{fig:ccc3} for $CCC_3$ is a surface of this type.

We may assign the cycles $c_x$ a single color class. When $n$ is even we may assign the cycles $e_{x,i}$ two color classes: one containing cycles of the form $e_{x,i}$ for even values of $i$ and the other containing cycles of the form $e_{x,i}$ for odd values of $i$. This three-coloring of $C$ (for even $n$) is an embedding of $CCC_n$ onto an $xyz$ surface, and therefore, by Theorem~\ref{thm:xyz surface}, $CCC_n$ is an $xyz$ graph.
\end{proof}

We have not determined whether the cube-connected cycles of odd order greater than three may be an $xyz$ graph.

\begin{figure}[t]
\centering
\includegraphics[height=1.8in]{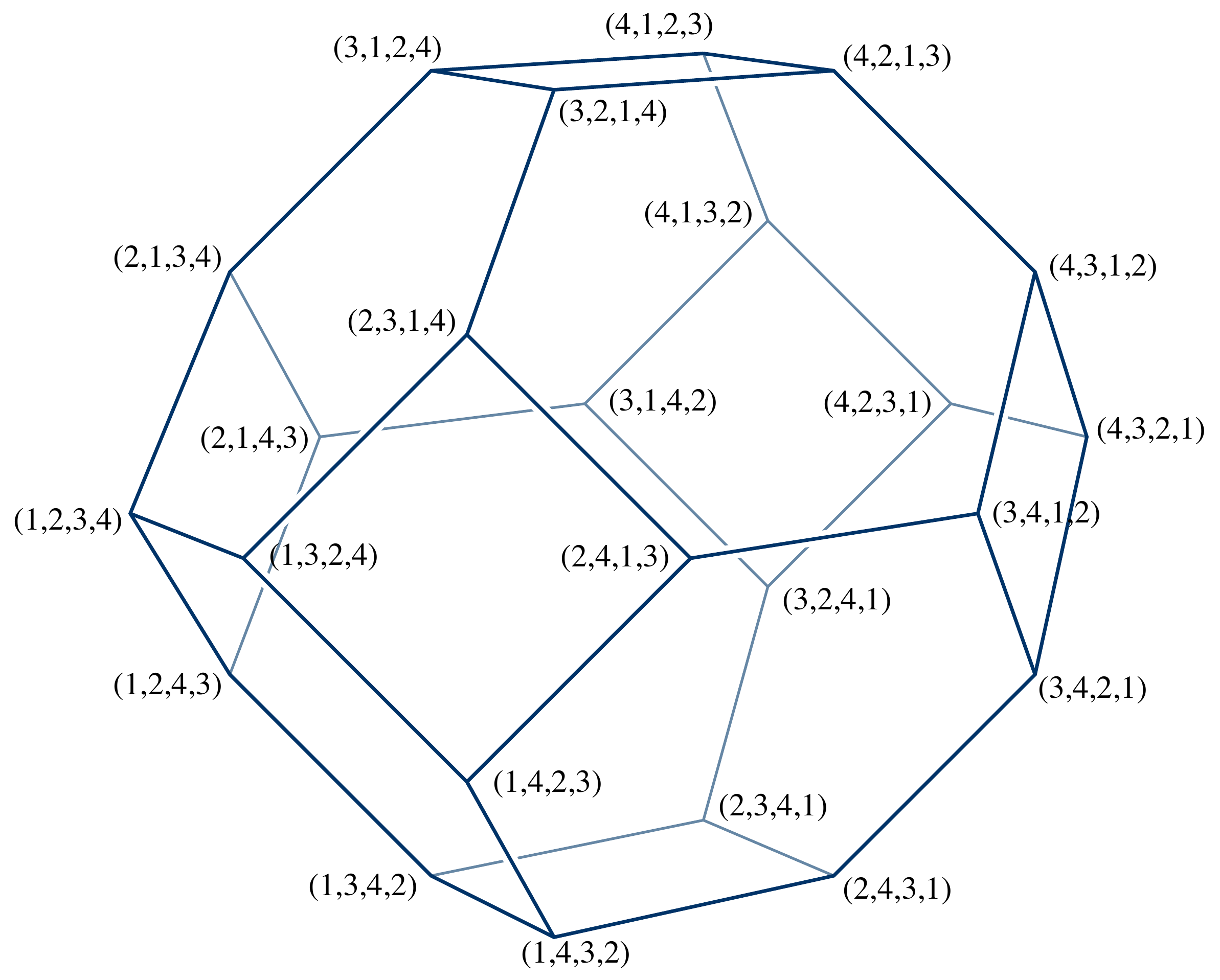}\qquad
\includegraphics[height=1.65in]{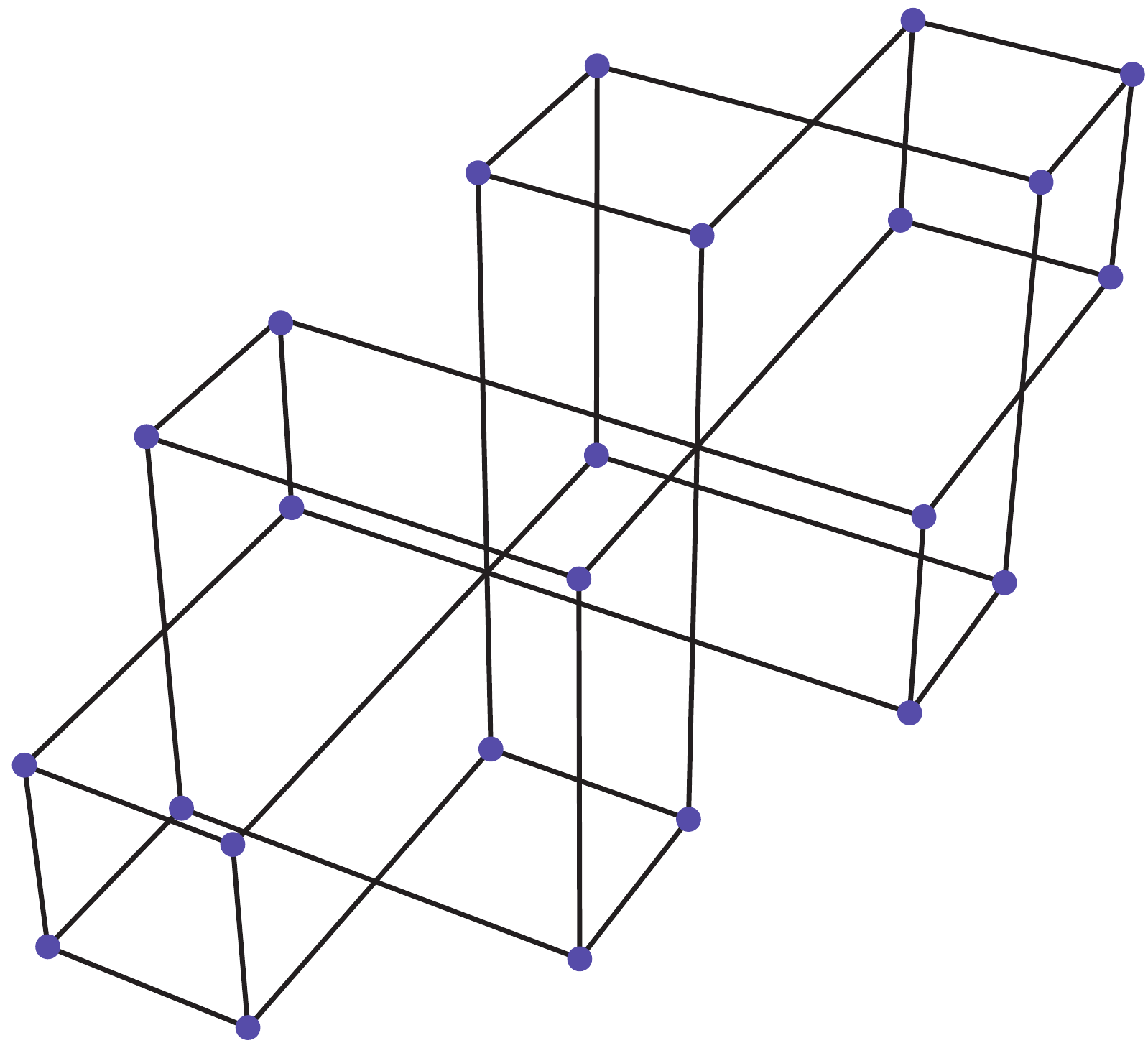}\qquad
\includegraphics[height=1.55in]{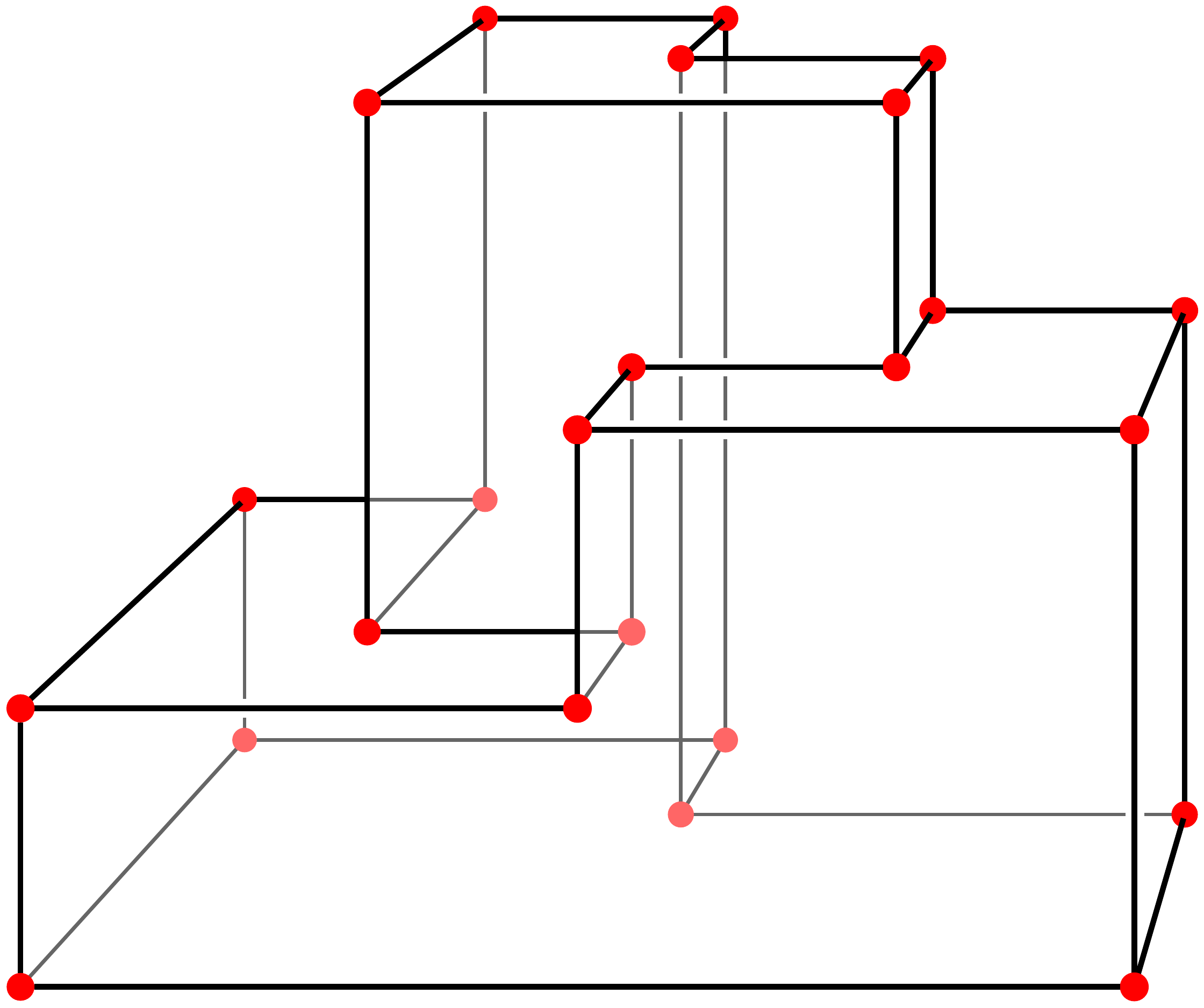}
\caption{The permutohedron (left) and two $xyz$ drawings of the corresponding Cayley graph (center and right). The center drawing is formed by connecting pairs of permutohedron vertices that differ by swapping consecutive coordinates and affinely transforming so that the edges are perpendicular; the right drawing permutes the values of each coordinate to realize the Cayley graph as the skeleton of an orthogonal polyhedron.}
\label{fig:permutohedron}
\end{figure}

Another important cubic Cayley graph is that of the \emph{symmetric group} of permutations on four elements, generated by transpositions of adjacent elements. This graph is isomorphic as an unlabeled graph to the skeleton \emph{permutohedron}, the convex hull of the 24 points formed by permuting the coordinates $(1,2,3,4)$, lying in the three-dimensional subspace $x+y+z+w=10$ of $\R^4$~\cite{GaiGup-SJAM-77,Pol-MISH-90}; however, the labeling of vertices of the Cayley graph and permutohedron differs. If we place the 24 permutations at the corresponding vertices of the permutohedron, but connect them by a different set of edges, the ones defined by the Cayley graph, then the edges fall into three parallel classes (corresponding to the three Cayley graph generators). If we then transform the drawing affinely so that these three classes are perpendicular, the result is an $xyz$ graph. Figure~\ref{fig:permutohedron} shows the permutohedron, the resulting $xyz$ drawing of its skeleton, and another $xyz$ drawing in which we have permuted the coordinate values manually in order to reduce the number of crossings. A different Cayley graph for the same symmetric group, generated by the permutations $(12)(3)(4)$, $(13)(2)(4)$, and $(14)(2)(3)$, is the 24-vertex symmetric graph shown later in Figure~\ref{fig:xyz-rhombs}.

\begin{figure}[t]
\centering\includegraphics[height=3in]{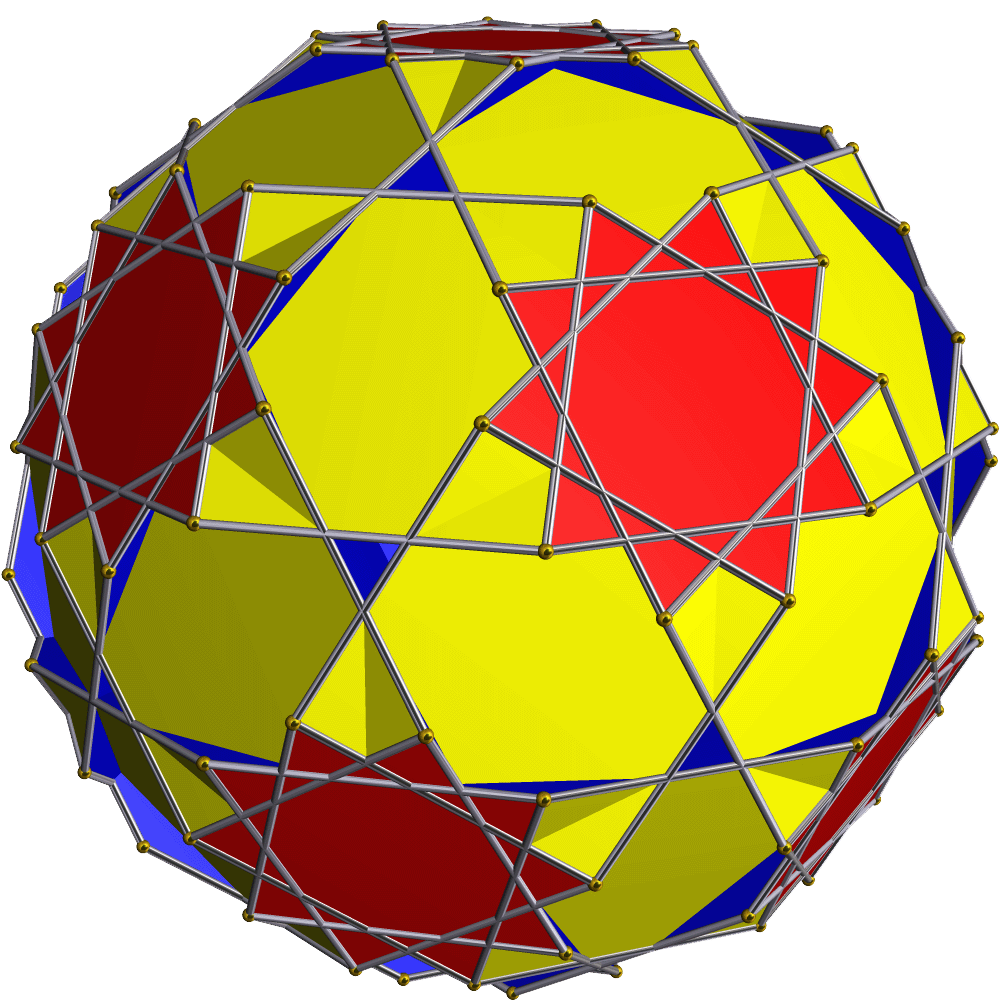}
\caption{The truncated dodecadodecahedron, from Wikimedia Commons, originally uploaded to Wikipedia by Tom Ruen in October 2005 and created using Robert Webb's Great Stella software (http://www.software3d.com/Stella.html). The vertices and edges of this shape form a Cayley graph for the symmetric group $S_5$, with generators $(12)(3)(4)(5)$ and $(1)(2345)$; the faces in the figure are 3-colored, giving an $xyz$ surface representation of the graph.}
\label{fig:trunc1212}
\end{figure}

Higher dimensional permutohedra have too many edges per vertex to be xyz graphs, but a different 
Cayley graph for the symmetric group $S_n$ is an $xyz$ graph whenever $n$ is an odd number greater than three. For any $n$, the symmetric group $S_n$ may be generated by two generators: the permutation that swaps the first two elements of an $n$-element sequence while leaving the remaining elements fixed, and the permutation that rotates the last $n-1$ elements of the sequence while leaving the first element fixed. For $n=3$ this just gives a hexagon (not an $xyz$ graph because each vertex only has two edges), and for $n=4$ it gives the truncated octahedron (isomorphic as a graph though not as a Cayley graph to the cube-connected cycles). Thus, the first case in which it gives an $xyz$ graph is $n = 5$: this Cayley graph for $S_n$ can be represented geometrically as the skeleton of a uniform polyhedron, the \emph{truncated dodecadodecahedron} (Figure~\ref{fig:trunc1212}), which has 30 square faces, 12 decagonal faces, and 12 star-shaped faces with ten vertices per face, interpenetrating each other to form a complex surface. The 3-face-coloring by which the truncated dodecadodecahedron can be recognized as an xyz surface coincides with the partition of its faces into different shapes.

To see that this Cayley graph for larger odd $n$, we need to examine the structure of Cayley $xyz$ graphs more generally.
A cubic Cayley graph may be formed from either of two types of generator set: three generators that are each their own group inverse elements, or two generators one of which is self-inverse and one of which is not. For instance, the permutohedron and the 24-vertex symmetric graph are formed from three self-inverse generators, while the cube-connected cycles and the truncated dodecadodecahedron have one self-inverse and one non-self-inverse generator. For a Cayley graph with three self-inverse generators, there is a natural partition of the edges into three matchings, where each matching consists of the edges formed by one of the generators; one may use Corollary~\ref{cor:test xyz graph} to determine whether this partition forms an $xyz$ graph. For a Cayley graph with one self-inverse and one non-self-inverse generator, it is not so obvious how to determine a partition into matchings from the Cayley graph, but there does exist a natural set of faces formed by the generators: if $a$ is the self-inverse generator and $b$ is the other generator, then one can form faces of the form $b^j$ and $(ab)^k$. That is, the faces of the form $b^k$ follow only the edges formed by generator $b$, repeating until the starting group element is reached; the faces of the form $(ab)^k$ alternate between edges formed by the two generators, and always follow the $b$-edges in the direction from a group element $e$ to the element $be$, again repeating until the starting group element is reached. Every edge of the graph belongs to two different faces: the $b$-edges belong to both an $(ab)^k$ face and a $b^j$ face, while the $a$-edges belong to two oppositely-oriented $(ab)^k$ faces. However, the result may not be an $xyz$ surface: two $(ab)^k$ faces may have more than one edge in common, or it may not be possible to three-color the faces. One may test using Corollary~\ref{cor:test xyz graph} whether any set of faces formed in this way provides a representation of the Cayley graph as an $xyz$ surface.

In the case of the two generators for $S_n$, when $n$ is odd, the product $ab$ of the two generators is the permutation that rotates the whole $n$-element sequence cyclically by one position. No two cycles of the form $(ab)^n$ can share more than one edge, and these cycles may be assigned two different colors according to whether the permutation that is cyclically shifted by the cycle is odd or even. The third face color may be assigned to the cycles $b^{n-1}$. Thus, the Cayley graphs of this type generate $xyz$-surface representations of $S_n$ for any odd~$n$.

\section{Symmetric graphs}

\begin{figure}[t]
\centering
\includegraphics[scale=1]{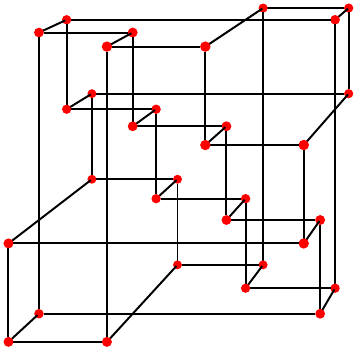}\quad
\includegraphics[scale=1]{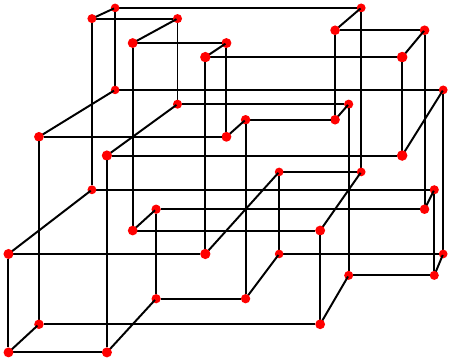}\quad
\caption{The Dyck graph  $F_{32}$ and the double cover of the dodecahedron $F_{40}$, two cubic symmetric graphs drawn as $xyz$ graphs.}
\label{fig:f32f40}
\end{figure}

\begin{figure}[t]
\centering
\includegraphics[height=2.5in]{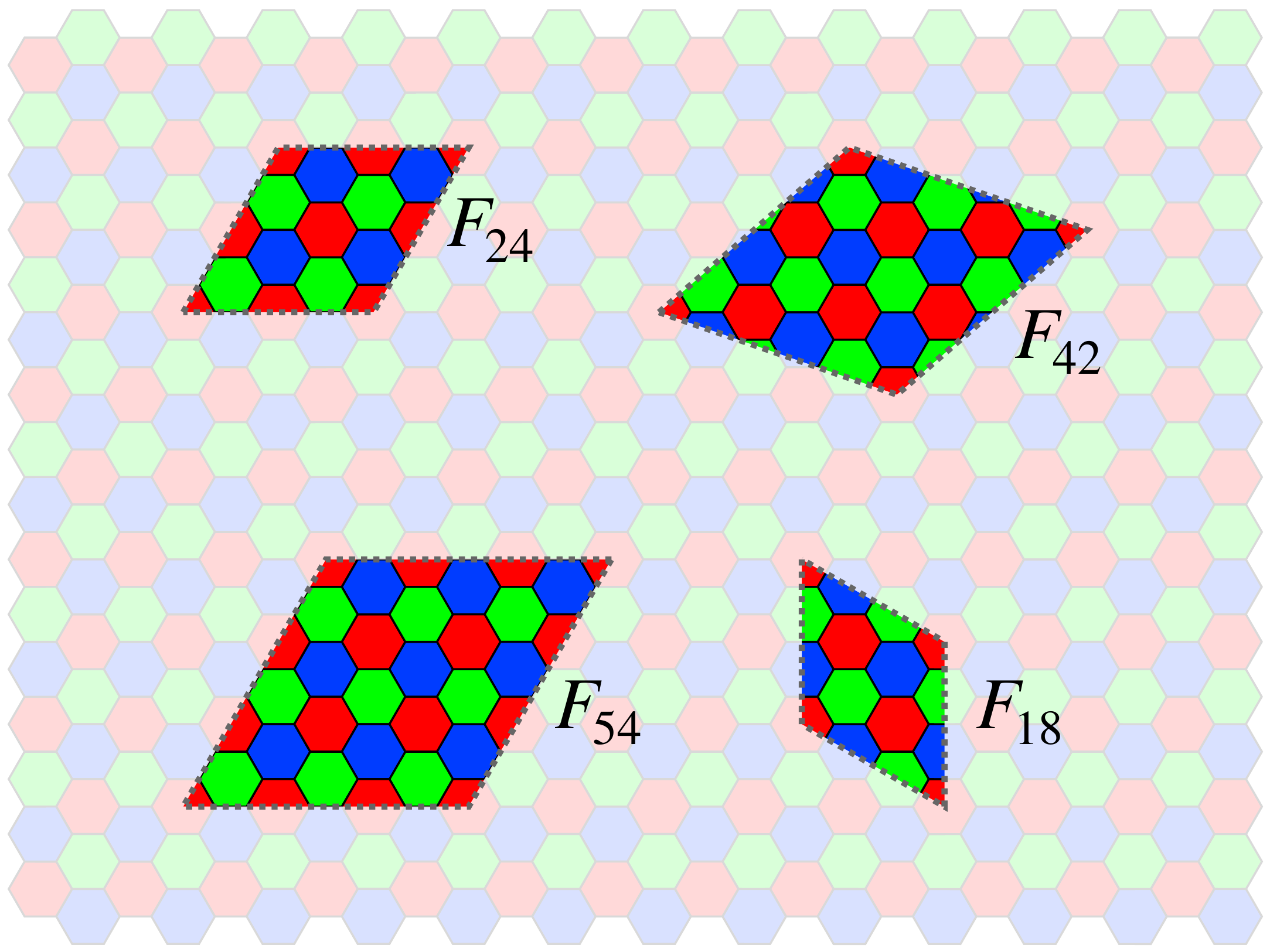}
\caption{Construction of cubic $xyz$ surfaces as toric quotients of the three-colored hexagonal tiling.}
\label{fig:xyz-rhombs}
\end{figure}

Consider the set of points $(x,y,z)$ in the $k\times k\times k$ grid for which $x+y+z$ is 0 or 1 (mod $k$). For $k=3$ this produces an embedding of the Pappus graph onto the torus; this embedding is combinatorially equivalent to, but geometrically different from, the graph shown in the right of Figure~\ref{fig:xyz333} and the left of Figure~\ref{fig:3x-tori}. (A very similar embedding is given by Pisanski~\cite{Pis-NZJM-07}.) Each graph formed in this way for arbitrary $k\ge 3$ is a \emph{symmetric graph}; that is, it has a group of symmetries that acts transitively on incident vertex-edge pairs. The graph $F_{32}$ shown on the left of Figure~\ref{fig:f32f40} is the \emph{Dyck graph}, a 32-vertex symmetric cubic graph embedded by the same construction with $k=4$. (Here $F_n$ refers to the unique $n$-vertex cubic symmetric graph as listed in the Foster census~\cite{Foster}.) Visible near the equatorial plane of the Dyck graph drawing are a number of six-vertex cycles that are not faces of the corresponding $xyz$ surface (they use edges parallel to all three coordinate axes, while the surface faces are restricted to axis-parallel planes); this pattern persists for larger $k$, and if one analogously forms an infinite $xyz$ graph from the points in a three-dimensional grid with coordinates summing to 0 or 1, the result is isomorphic as a graph to the hexagonal tiling of the plane~\cite{Epp-GD-08-ids}.

A different construction for cubic symmetric $xyz$ graphs is possible, based on the infinite tiling of the plane by regular hexagons. Three-color the hexagons of this tiling, choose a rhombus with angles of $\pi/3$ and $2\pi/3$, having its vertices at the centers of tiles that are all the same color, and form a torus by gluing opposite sides of this rhombus together. The result, as shown in Figure~\ref{fig:xyz-rhombs}, is an $xyz$ surface. The graph embedded on this surface is symmetric, because we can transform any incident vertex-edge pair into any other such pair by a combination of translations and rotations by an angle of $\pi/3$. Note that this construction does not always produce a \emph{regular map}, having symmetries taking any flag (incident triple of vertex, edge, and face) to any other flag; for instance, the surface for $F_{42}$ shown in the upper right of the figure is not regular. The graph $F_{24}$ shown in the upper left of the figure is another Cayley graph for the symmetric group on four elements, generated by the three permutations $(12)$, $(13)$, and $(14)$.

When $n=18q^2$ for some $q$, one can form an $n$-vertex symmetric graph using both of the constructions above, either by forming a torus from a rhombus containing $n/2$ hexagons, with sides parallel to the edges of the hexagonal tiling, or by using the set of points congruent to 0 or 1 in a $3q\times 3q\times 3q$ grid. Both graphs formed in this way are isomorphic, as may be seen using the equatorial hexagons described above for the $k\times k\times k$ grid graph, but (except for $k=1$) the $xyz$ graph embeddings resulting from these constructions are inequivalent: the $xyz$ surface resulting from the $k\times k\times k$ has fewer faces with more vertices per face than the torus formed from a rhombus in the hexagonal tiling. For instance the 72-vertex cubic symmetric graph $F_{72}$ may be represented as an $xyz$ surface with 18 12-vertex faces (the points congruent to 0 or 1 mod 6 in a $6\times 6\times 6$ grid) or with 36 6-vertex faces (a rhombus containing 36 hexagons).

Figure~\ref{fig:f32f40}, right, shows another cubic symmetric graph, $F_{40}$, that does not fit into either of these constructions. $F_{40}$ is the double cover of the regular dodecahedron; that is, it is the bipartite graph formed by making two copies of each dodecahedron vertex, colored black and white, and connecting the white copy of each vertex to the black copy of each of its neighbors. Its $xyz$ graph representation has faces of three types: two decagons formed as the double covers of a pair of opposite dodecahedron faces, two more decagons formed from the double cover of the equator between those two faces, and ten octagons formed as the boundary of a pair of adjacent dodecahedron faces that lie on opposite sides of the equator. There are six ways of choosing two opposite faces from which the decagons are formed, and once that choice is made there remain two ways of choosing the octagons to form an $xyz$ surface, so $F_{40}$, viewed as a labeled graph, has 12 combinatorially distinct $xyz$ surface representations.

We applied our implementation of an $xyz$ graph embedding algorithm to the Foster census of symmetric cubic graphs~\cite{Foster} and did not find any other $xyz$ graphs of this type on 56 or fewer vertices.

\section{Planar graphs}

We may exactly characterize the planar $xyz$ graphs.

\begin{lemma}
\label{lem:planar uniqueness}
Let $G$ be a planar graph, and $C$ be the family of face cycles of a planar embedding of $G$. Then any $xyz$ surface representation of $G$ must have $C$ as its face set.
\end{lemma}

\begin{proof}
Suppose for a contradiction that there is an $xyz$ surface representation of $G$ with a different face set $C'$. Since every edge of $G$ must remain covered twice by cycles in $C'$, $C'$ must contain a face $f$ that does not belong to $C$. By the Jordan curve theorem, $f$ separates the plane in which $G$ is embedded with face set $C$ into an inside and an outside, each of which must contain edges or vertices of $G$ since $f$ is not itself a face of $C$. Let $e$ be an edge of $f$ that connects an interior component of the remaining graph with an exterior component, and consider the cycle $f'$ in $C'$ that contains $e$ but is not $f$. $f'$ contains a path connecting the endpoints of $e$ that starts interior to $f$ and ends exterior to $f$; by the Jordan curve theorem this path must cross $f$, and (as $G$ must be cubic to have an $xyz$ surface representation) when it crosses $f$ it must do so by containing another edge of $f$. Thus, $f$ and $f'$ share at least two edges, violating the requirement of $xyz$ surfaces that each pair of faces share at most a single edge and contradicting the assumption that $C'$ is the face set of an $xyz$ surface representation of $G$.
\end{proof}

\begin{theorem}
\label{thm:planar}
Let $G$ be a planar graph. Then $G$ is an $xyz$ graph if and only if $G$ is bipartite, cubic, and 3-connected. If it is an $xyz$ graph it has a unique representation as an $xyz$ surface, up to permutation of the face colors of the surface.
\end{theorem}

\begin{proof}
If $G$ is a planar $xyz$ graph, it must be cubic, and by Theorem~\ref{thm:3con} it must be 3-connected. By Lemma~\ref{lem:planar uniqueness} its corresponding $xyz$ surface is unique and must be topologically a sphere, so by Theorem~\ref{thm:orientability} it must be bipartite.

Conversely, suppose that $G$ is a bipartite cubic 3-connected planar graph. By Steinitz' theorem $G$ can be represented as the skeleton of a convex polyhedron; by convexity, every pair of faces meets in at most a single edge. Heawood~\cite{Hea-QJPAM-98} proved that the faces of any bipartite convex polyhedron can be 3-colored and this coloring provides an $xyz$ surface representation of $G$. By Theorem~\ref{thm:xyz surface}, $G$ is an $xyz$ graph.
\end{proof}

\begin{corollary}
We may test in linear time whether a given planar graph $G$ is an $xyz$ graph.
\end{corollary}

The bipartite 3-connected cubic planar graphs include all known cubic partial cubes~\cite{Epp-EJC-06} with a single exception, the Desargues graph, which is a cubic partial cube but is not planar and turns out not to be an $xyz$ graph.

\section{Nonplanar graphs}

We have seen already a construction of cubic symmetric $xyz$ graphs, in which we form a graph from
 the set of $2k^2$ points $(x,y,z)$ in the $k\times k\times k$ grid for which $x+y+z$ is 0 or 1 (mod $k$). 
 This graph has $3k^2$ vertices, but only $3k$ faces (one for each axis-aligned plane) so the Euler characteristic of the $xyz$ graph formed in this way is $3k-k^2$.

If $G$ and $G'$ are $xyz$ graphs, with designated vertices $v$ and $v'$, we may form the \emph{connected sum} of $G$ and $G'$ by aligning the two graphs in $\R^3$ so that $v$ and $v'$ coincide (and so that no pairs of vertices, one from $G$ and one from $G'$, lie on an axis-parallel line unless both vertices in the pair are adjacent to $v$ and $v'$) and then by removing $v$ and $v'$, leaving in their place a non-vertex point where the lines through three edges cross. The 14-vertex planar graph in the center of Figure~\ref{fig:xyz333} can be viewed in this way as a connected sum of two cubes. In terms of $xyz$ surfaces, the connected sum operation can be viewed as cutting the two surfaces by a small disk surrounding each of $v$ and $v'$, and gluing the three faces surrounding this hole on one surface to the faces of corresponding colors surrounding the hole on the other surface, to form a handle connecting the two surfaces. By forming connected sums of tori and projective planes (the $xyz$ graphs on the left and right of Figure~\ref{fig:xyz333} respectively), we may form $xyz$ surfaces of any topological type.

\begin{figure}[t]
\centering
\includegraphics[width=4in]{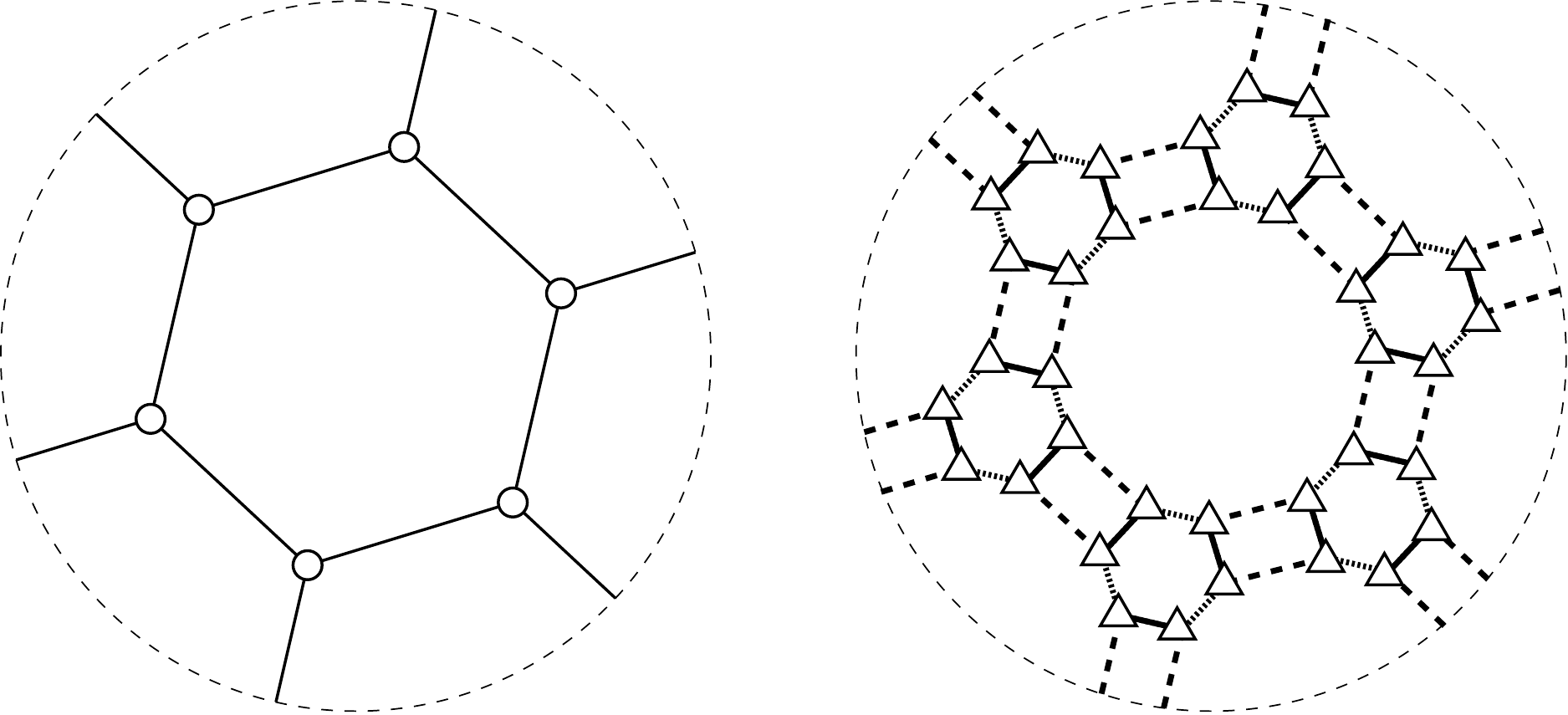}
\caption{An embedding of $K_{3,3}$ in the projective plane with one hexagonal face and three quadrilateral faces (left) and a GEM representation of the embedding (right). From~\cite{Epp-SODA-03}.}
\label{fig:gem}
\end{figure}
 
There is an alternative construction which also allows arbitrary surfaces to be represented as $xyz$ surfaces: the \emph{graph encoded map} (GEM)~\cite{BonLit-95,Epp-SODA-03}. Let $G$ be any graph embedded on a 2-manifold in such a way that each face of the embedding is a topological disk bounded by a simple cycle of $G$. A \emph{flag} of this embedding is a triple of a vertex, edge, and face that are all incident to each other, and the graph encoded map $M$ of this embedding is a 3-edge-colored cubic graph, having a vertex for each flag of the embedding of $G$. Two vertices of $M$ are adjacent if the corresponding two flags differ only in a vertex, differ only in an edge, or differ only in a face; the edge coloring of $M$ determines which type of difference each edge of $M$ represents. $M$ itself can be embedded on the same surface, with a $2k$-cycle for each vertex of degree $k$ in $G$ or each face in $G$ that is surrounded by $k$ edges, and a $4$-cycle for each edge of $G$. These cycles form an $xyz$ surface, in which the color of a face in the GEM is determined by whether it represents a vertex, face, or edge in $G$, so $M$ is an $xyz$ graph. Figures~\ref{fig:gem} depicts an embedded graph and its GEM;  Figure~\ref{fig:TuckerGroup} can also be interpreted as a GEM, for an embedding of the M\"obius--Kantor graph on the double torus. GEMs can be distinguished among other $xyz$ surfaces by the property that one of the three color classes of faces (the set of faces of the GEM that represents edges in $G$) consists only of quadrilaterals; any $xyz$ surface with this property can be interpreted as a GEM of two dual graphs or multigraphs on the same surface, where the duality between the graphs is represented by exchanging the other two color classes of the $xyz$ surface coloring.

\begin{figure}[t]
\centering\includegraphics[width=6in]{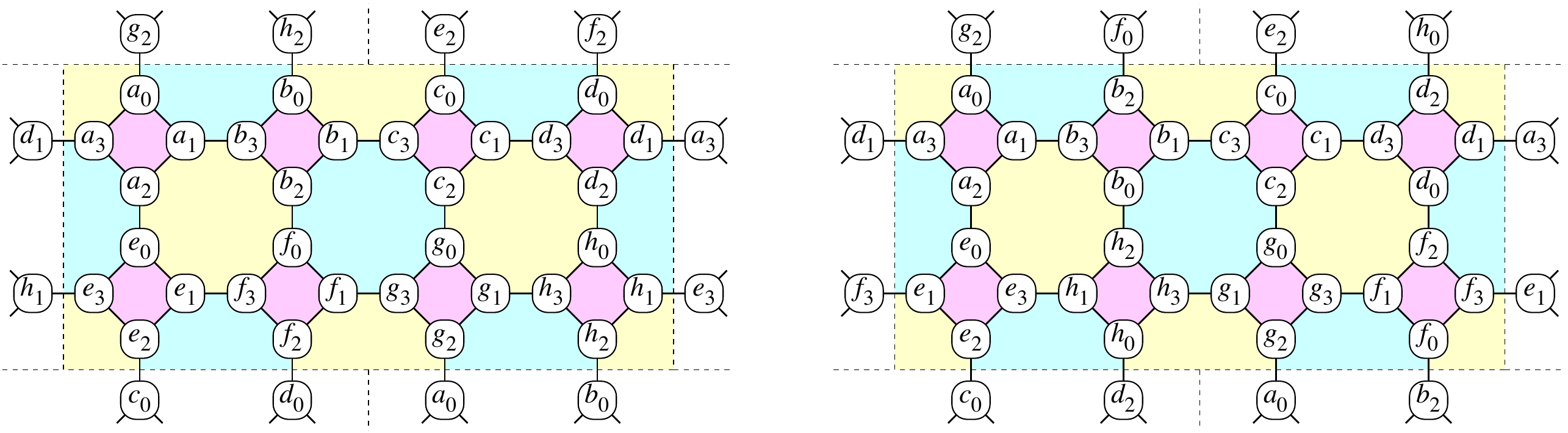}
\caption{A graph that can be embedded on the torus as an $xyz$ surface in two different ways.}
\label{fig:ambig}
\end{figure}

The 32-vertex graph shown in Figure~\ref{fig:ambig} consists of eight ``diamonds'' of four vertices each, connected in pairs to form a cubic graph. The central colored rectangle of the graph denotes a surface that may be glued to itself along its boundary to form a torus; the left and right ends of the rectangle are connected to each other in the usual way, but the top and bottom ends are connected after a shift to form a brick wall pattern. Several vertices are drawn twice, both inside the rectangle and a second time outside it, to show the edges that cross these glued rectangle boundaries. The faces of the surface (eight diamonds and eight octagons) are colored, showing that it forms an $xyz$ surface. One way to construct this embedding of this graph is to form a torus by gluing together the opposite edges of a $2\times 2$ grid of squares, and then to form the GEM of the resulting map.
However, the same graph may be drawn in two different ways, as shown in the figure. The left and right sides of the figure show two embeddings that are isomorphic as unlabeled graphs but nonisomorphic when the graph vertices are labeled, as they are in the figure, although they both have the same sets of vertices and edges. The octagonal faces on the right-hand embedding form zigzag paths through the left-hand embedding: the blue faces correspond to cycles in the left embedding that zigzag from top left to bottom right, while the yellow faces in the right drawing zigzag from the bottom left to the top right of the left drawing. This second embedding is related to the first one by applying a half-twist to each of the edges connecting pairs of pink diamonds. Thus, the graph in the figure is an $xyz$ graph in two different ways, showing that the uniqueness of $xyz$ surface representations for planar graphs does not directly generalize to other surfaces.

The same brick wall construction works with larger $4k\times2k$ arrays of diamonds. In each case there is an obvious torus embedding and a less obvious embedding formed by zigzag paths. But for the larger graphs, the two embeddings are not isomorphic: the zigzag paths are longer and fewer in number, so the nonobvious embeddings have higher genus. Graphs with arbitrarily large numbers of distinct embeddings or even arbitrarily large numbers of embeddings with different genuses are also easily constructed by connected sums of these graphs.

This ambiguously-embeddable graph will play a key role in our NP-completeness proof in Section~\ref{sec:complexity}.

\section{Covering}

A \emph{covering} of a graph $G$ consists of a graph $\hat G$, and a map $f$ from vertices of $\hat G$ to vertices of $G$, such that (1) for each edge $\hat u\hat v$ in $H$, $f(\hat u)f(\hat v)$ is an edge in $G$, and (2) for each edge $uv$ in $G$ and each vertex $\hat u$ in $f^{-1}(u)$ there is a unique neighbor $\hat v$ of $\hat u$ with $\hat v in f^{-1}(v)$.  A \emph{covering} of a map $M$ consists of a map $\hat M$ and a continuous function $f$ from $\hat M$ to $M$ that maps vertices to vertices, edges to edges, and faces to faces, in such a way that the restriction of $f$ to the underlying graph of $M$ is a covering. A covering has \emph{finite ply} if the inverse image of every point in $G$ or $M$ is finite; if $G$ or $M$ is connected and the covering has finite ply, then the inverse image of every point on a vertex or edge of the graph has the same cardinality, known as the \emph{ply} of the covering. A covering is \emph{unbranched} if every point of $\hat M$ has a neighborhood within which $f$ is a homeomorphism, and \emph{branched} otherwise; in an unbranched covering, the number of edges of a face is the same in $\hat M$ as in its image in $M$, while in a branched covering, faces of $\hat M$ may correspond to faces with fewer edges in $M$.

We have seen that not every graph is an $xyz$ graph and not every map is an $xyz$ surface. But, as we now show, every cubic graph may be covered by an $xyz$ graph, and more generally that every cubic map may be covered by an $xyz$ surface. To begin the proof, recall that a map is \emph{polyhedral} if two faces of the map cannot intersect in more than a single edge.

\begin{lemma}
\label{lem:polyhedral-cover}
Any cubic map without self-loops may be covered by a polyhedral cubic map in which every face has even length.
\end{lemma}

\begin{proof}
Define a group $G=\Z_2^m$, where $m$ is the number of edges in the given map, and label each edge in the map by a generator of $G$. Form a new map $M\times G$ that has one vertex or edge for each pair of a vertex or edge of $M$ and an element of the group $G$; if $v$ is a vertex of $M$, and $vw$ is an edge of $m$ labeled by a group element $e$, then the map $M\times G$ includes an edge from each vertex $(v,x)$ to the vertex $(w,ex)$. Form the faces of the map $M\times G$ as follows: if $C$ is the cycle of edges forming a face of $M$, $v$ is a vertex of $M$, and $(v,x)$ is a vertex of $M\times G$, then form a face in $M\times G$ that repeatedly follows edges in $M\times G$ corresponding to the edges of $C$, starting from $(v,x)$, until it reaches $(v,x)$ again.

Then $M\times G$ is, by construction, a map that covers $M$. Every face in $M\times G$ has twice as many vertices as the corresponding face in $M$: traversing a cycle $C$ once produces an element of $G$ that is the product of all the generators labeling the edges of $C$, and traversing the cycle again cancels each of these generators. No two faces may have more than a single edge in common, for otherwise the two paths within the two faces connecting two minimally-distant shared edges would consist only of edges used an even number of times within each path, an impossibility in a cubic map.
\end{proof}

The ply of this covering is large ($2^m$), but may be reduced by replacing the edge labels by randomly chosen elements of $\Z_2^k$, where $k$ is proportional to the logarithm of the number of edges in the largest face of $M$, and applying the Lov\'asz Local Lemma. It may be necessary to use ply at least proportional to the number of vertices in $M$, as may be seen by considering the oriented map with one $n$-gon and one $2n$-gon (where $n$ is a multiple of
four) formed by adding $n/2$ edges connecting opposite pairs of vertices
of the $n$-gon to form a M\"obius ladder. In this map, the $2n$-gon has $n$ adjacencies to
itself (on both sides of the $n/2$ diagonal edges). If $f$ is a face
corresponding to the $2n$-gon in any polyhedral cover of $M$, the corresponding set of
adjacencies must connect it to at least $n$ different copies of itself.

Define a \emph{perfect face cover} of a map to be a set of faces that meets each vertex exactly
once. If a cubic map is 3-face-colorable, each of its three color classes forms
a perfect face cover. A perfect face cover of a connected map, if it exists, may be found in linear time:
form a graph $H$ that has a vertex for each face of the map $M$, and connect two vertices $f_1$ and $f_2$ of $H$ by an edge whenever there exists an edge $uv$ of $M$ where $u\in f_1$, $v\in f_2$, and $uv$ belongs neither to $f_1$ nor $f_2$.  if a perfect face cover exists, it must correspond to a connected component of $H$.

\begin{theorem}
\label{thm:sixfold}
Any connected polyhedral cubic map $M$ in which every face has even length has an unbranched cover by a connected $xyz$ surface; the ply of the cover is at most six.
Specifically, if $M$ is already an $xyz$ surface, then the ply is one.
If $M$ is not an $xyz$ surface, but has a perfect face cover, then the ply is two.
If $M$ has no perfect face cover, but is orientable and has a bipartite underlying graph, then the ply is three. If $M$ has no perfect face cover, but is not orientable and every orientation-reversing cycle in the underlying graph has odd length, then the ply is again three. And in the remaining cases, the ply is six.
\end{theorem}

\begin{proof}
Form a sixfold cover $\hat M$ in which each vertex $v$ of $G$ is covered by six copies, one for each possible way of choosing distinct orientations in $\{x,y,z\}$ for each of the three incident edges at $v$.
Let each face $f$ of $G$ be covered again by six copies, one for each possible way of labeling the boundary by edges that alternate between two of the three orientations. These faces and vertices may be glued together to form a sixfold cover of $M$: there is only one way to glue the six copies of each of
the three faces incident to a vertex $v$ in such a way that each of the six
copies of $v$ gets edges of all three colors. This cover, however, may have more than one connected component. Due to the symmetry of the construction, the ply of the cover is exactly $6/c$ where $c$ is the number of connected components.

If $M$ is an $xyz$ surface itself, then this construction merely creates each possible 3-face-coloring of $M$, and there are six components for the six choices of coloring. Otherwise, there must be fewer than $c$ components.

Any $M$ is not an $xyz$ surface, but has a perfect face cover $C$, then we may 3-face-color a component of $\hat M$ by choosing one color for the faces in $C$ and using the two remaining colors for the remaining faces. Each choice of a color for $C$ must lead to a single connected component of $\hat M$, for otherwise we would have six components as can only happen when $M$ is an $xyz$ surface. Thus, in this case, there are three components.

If $M$ does not have a perfect face cover, then in every connected component of $\hat M$ every face of $M$ must have at least three preimages, one for each possible color that face may be given. However, $\hat M$ has two copies of each color of each face $f$ of $M$, one for each of the two ways that faces of the other two colors may alternate around $f$. If $\hat M$ includes a path that connects both copies of some color of $f$, it has a single connected component; if it does not include such a path, then it has two connected components. Such a path must correspond to an orientation-preserving odd-length cycle or an orientation-reversing even-length cycle. Thus, $\hat M$ has two components if and only if no such cycle exists. If $M$ is orientable, such a cycle exists if and only if the graph is nonbipartite. If $M$ is non-orientable, and there is an orientation-preserving odd-length cycle, then traversing it together with an orientation-reversing odd-length cycle will produce an orientation-reversing even-length cycle, so $\hat M$ has two components if and only if there is no orientation-reversing even-length cycle.
\end{proof}

This provides another proof of Heawood's theorem~\cite{Hea-QJPAM-98} that any planar polyhedral map with even-length faces may be 3-face-colored: a sphere can have no nontrivial cover, so the map whose existence is guaranteed by the theorem must fall into the ply-one case.

\begin{corollary}
Any cubic map has a cover by an $xyz$ surface, and any cubic graph has a cover by an $xyz$ graph.
\end{corollary}

\section{Complexity of $xyz$ graph recognition}
\label{sec:complexity}

\begin{figure}[t]
\centering\includegraphics[width=3in]{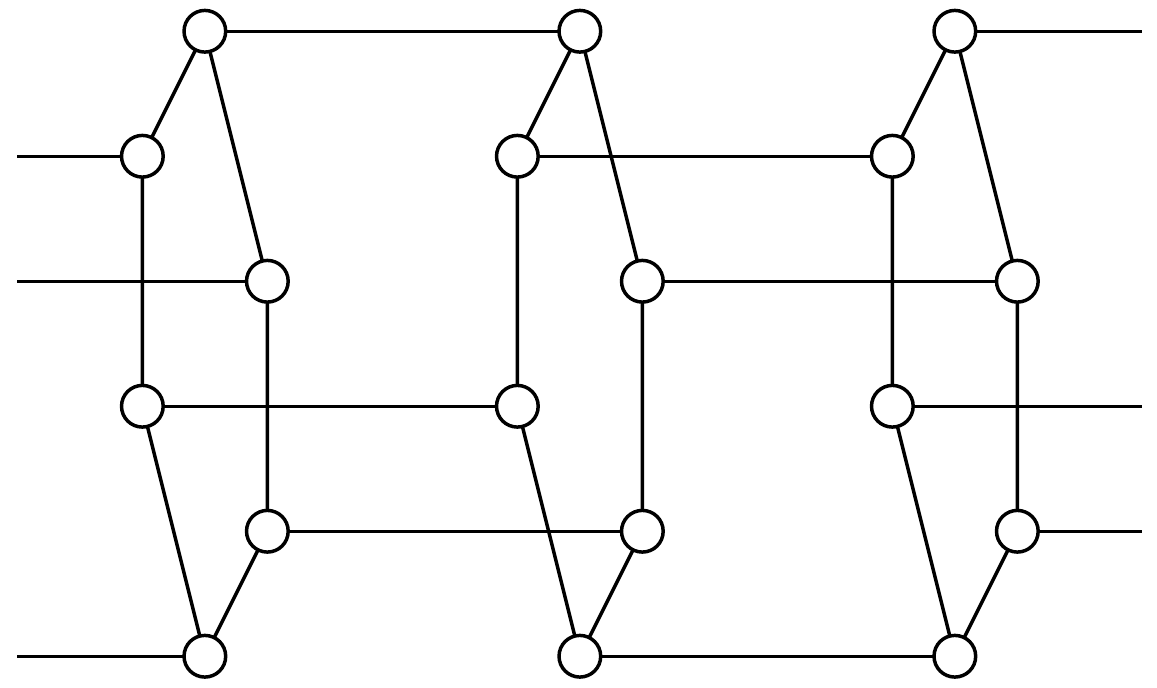}
\caption{The connector gadget, part of an NP-completeness proof for $xyz$ graph recognition.}
\label{fig:connector}
\end{figure}

We will show that recognizing $xyz$ graphs is NP-complete, via a reduction from graph 3-colorability. Throughout this section, when we refer to the \emph{orientation} of an edge, we mean the choice of which coordinate axis to make the edge parallel to in an $xyz$ graph representation; this choice will correspond to the choice of a color in our 3-colorability reduction. Our reduction will connect together gadgets (certain cubic graphs) via connected sums. The key gadgets are the ambiguously embeddable torus from Figure~\ref{fig:ambig}, and another part of a cubic graph shown in Figure~\ref{fig:connector}, which we call the \emph{connector gadget}.

The connector gadget, as drawn in the figure, can be viewed as part of a cylindrical piece of surface, in which the gadget partitions the surface of the cylinder into rings of three curved hexagonal faces; each such ring can be 3-colored. As we now show, that is the only possible way in which this gadget can be part of an $xyz$ surface.

\begin{figure}[t]
\centering\includegraphics[width=4in]{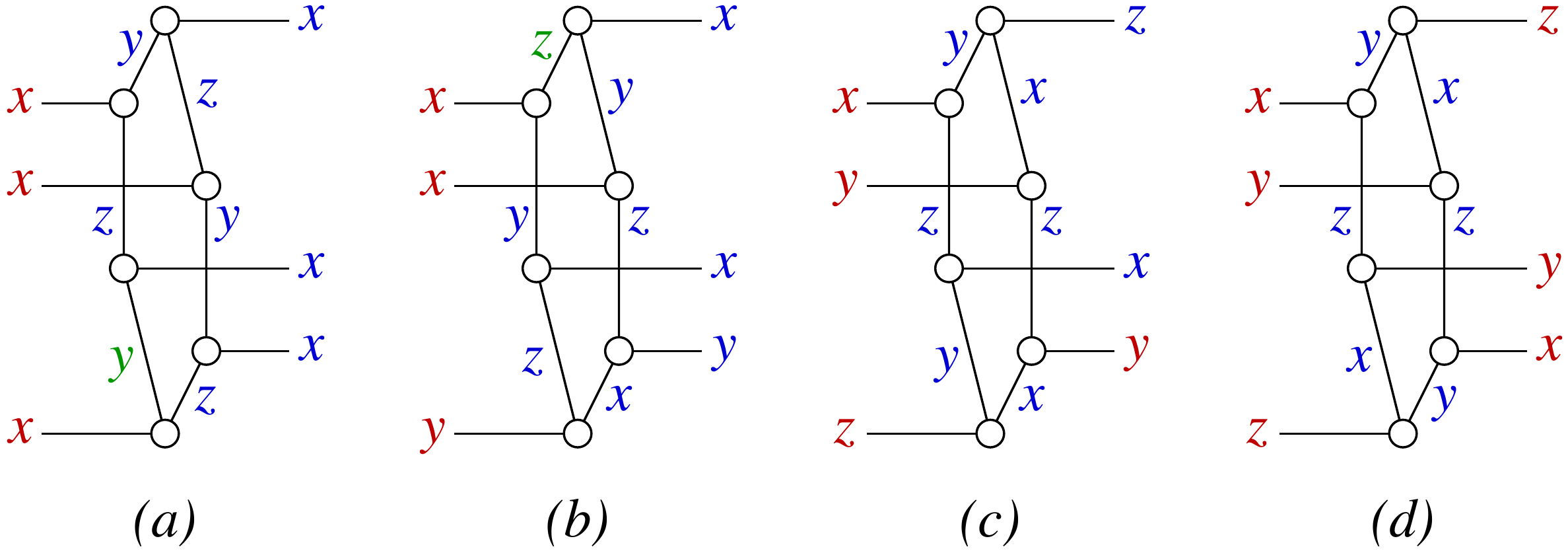}
\caption{Case analysis for Lemma~\ref{lem:connector}. In each case the red letters denote the edge orientations that specify that case, the green letter (if present) denotes an arbitrary choice of orientation that may be made once the red orientations are set, and the blue letters denote edge orientations that are forced by the red and green orientations and the requirement that incident edges be perpendicular.}
\label{fig:concases}
\end{figure}

\begin{lemma}
\label{lem:connector}
Suppose that the connector gadget shown in Figure~\ref{fig:connector} is a subgraph of some larger $xyz$ graph $G$. Then in any $xyz$ graph representation of $G$, the three edges entering the gadget from the left of the figure must be mutually perpendicular and lie on three lines that all meet in a single point. Similarly, the three edges entering the gadget from the right of the figure must be mutually perpendicular and lie on lines that all meet at a single point. If the left three lines are $A$, $B$, and $C$ (as shown top-down in the figure) and the right three lines are $D$, $E$, and $F$ (again, top-down), then $A$ and $F$ must be parallel, $B$ and $E$ must be parallel, and $C$ and $D$ must be parallel.
\end{lemma}

\begin{proof}
We examine cases, showing that in each case other than the one described the axis-parallel polygons of the drawing could not form an $xyz$ surface.

First, suppose that $A$, $B$, and $C$ are all parallel to each other; by symmetry we may assume that they are parallel to the $x$ axis, as shown by the red labels in Figure~\ref{fig:concases}(a). Then the hexagon to which these three edges attaches must have edges that alternate $yzyzyz$ in one of two ways (the green and blue labels in the same figure), and the next three edges drawn as horizontal in the drawing must also be parallel to the $x$ axis. Repeating this reasoning, all twelve horizontal edges in the drawing of Figure~\ref{fig:connector} must be parallel to the $x$ axis. There would be six paths through this drawing labeled $xyxyx$ or $xzxzx$, which must be part of axis-parallel polygons in any $xyz$ surface for this drawing. But these six paths share 12 $x$-parallel edges, while they can be at most part of three $xy$-parallel and three $xz$-parallel polygons, which can together share at most 9 edges in any $xyz$ surface. Therefore, if $A$, $B$, and $C$ are drawn parallel, the drawing cannot form an $xyz$ graph.

Next, suppose that two of $A$, $B$, and $C$ are parallel to each other, while the third is not.
We may assume by symmetry that two ($A$ and $B$) are parallel to the $x$ axis while the third ($C$) is parallel to the $y$ axis, as shown in Figure~\ref{fig:concases}(b). Then, if we pick arbitrarily the orientation of one of the two edges connecting $A$ and $B$, and then orient the remaining edges of the hexagon to which $A$, $B$, and $C$ attach by choosing the only remaining orientation for an edge whenever two adjacent edges have already been oriented in perpendicular directions, we find that the edges around this hexagon must be oriented in the pattern (clockwise starting from the endpoint of $C$) $xzyzyz$ or its reverse. This forces the orientation of the next three horizontal edges of the gadget to again have two edges parallel to the $x$ axis and the third parallel to the $z$ axis, and so forth from left to right across the gadget. There must then be two paths labeled $xyxyxyx$ that extend from left to right across the gadget, and one path labeled $xzxzxzx$ that also extends from left to right across the gadget; these three paths together share four $x$-parallel edges, but they belong to at most one $xz$-parallel polygon and at most two $xy$-parallel polygons that can only share a total of two edges in any $xyz$ surface. Therefore, in this case, we again fail to get an $xyz$ graph.

In the third case, $A$, $B$, and $C$ are labeled differently from each other (without loss of generality they are labeled $x$, $y$, and $z$ respectively, but the horizontal edge in the next layer of the gadget to the right that is farthest from $A$ is not labeled $x$; without loss of generality its label is $y$, as shown in Figure~\ref{fig:concases}(c). In this case, when we assign orientations to edges that are forced by having two perpendicular adjacent edges, as shown in the figure, we get two paths through the gadget starting from $C$ that are labeled $zxzxz$ and $zyzyz$, and that share both their starting and their ending edges; this would cause the corresponding two faces of any surface embedding to share two edges, impossible in an $xyz$ surface.

Finally, the only remaining case is that $A$, $B$, and $C$ are labeled distinctly, and that each label matches the label of the opposite edge in the next layer of horizontal edges, as shown in Figure~\ref{fig:concases}(d). $A$, $B$, and $C$ are mutually perpendicular, and form the boundaries of three mutually adjacent faces of the $xyz$ surface along paths through the gadget labeled $xyxy$, $yzyz$, and $zxzx$. Since the three lines through $A$, $B$, and $C$ lie on the intersections of the pairs of planes through these three faces, the three lines must meet at the point where these three planes meet.
\end{proof} 

We will use this gadget in our NP-completeness reduction by attaching it to other gadgets via the connected sum construction: if $G$ is a (possibly disconnected) cubic graph with designated vertices $u$ and $v$, define a graph $G_{u,v}$ by deleting vertices $u$ and $v$ from $G$, reconnecting the edges that were incident to $u$ to the left side of a connector gadget, and reconnecting the edges that were incident to $v$ to the right side of a connector gadget.

\begin{figure}[t]
\centering\includegraphics[width=5.5in]{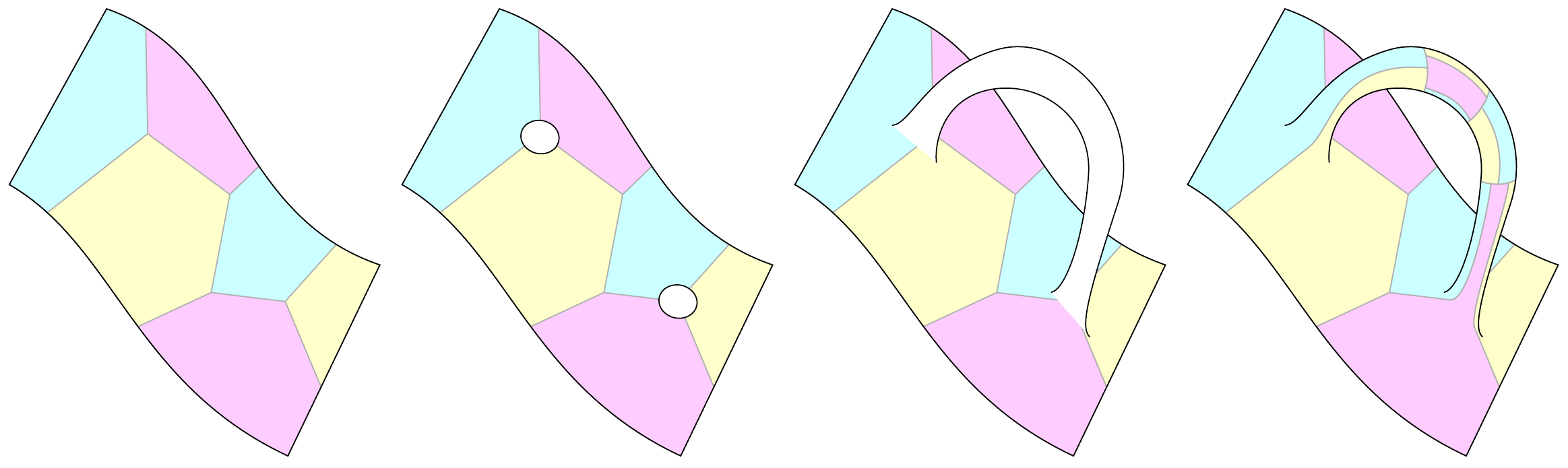}
\caption{Creating an $xyz$ surface for $G_{u,v}$ from an $xyz$ surface for $G$.}
\label{fig:contop}
\end{figure}

\begin{lemma}
\label{lem:abstract attachment}
The orientations of edges of $G_{u,v}$ that come from valid $xyz$ graph representations are in one-to-one correspondence with the orientations of edges of $G$ that come from valid $xyz$ graph representations and that assign consistent orientations to the edges of $G$ at $u$ and at $v$ (where the notion of consistency is determined by the precise pattern of attachment of edges of $G$ to the connector gadget in the construction of $G_{u,v}$. In particular, $G_{u,v}$ is an $xyz$ graph if and only if $G$ has an $xyz$ graph representation in which $u$ and $v$ are consistently oriented.
\end{lemma}

\begin{proof}
From any $xyz$ graph representation of $G_{u,v}$, we may return to an $xyz$ graph representation of $G$ by deleting the connector gadget and replacing the deleted vertices of $G$ at the points of incidence of the lines that the connector gadget edges lie on; these points of incidence are guaranteed to exist by Lemma~\ref{lem:connector}. By Lemma~\ref{lem:connector}, the edges of the connector gadget in $G_{u,v}$ will be consistently oriented at $u$ and at $v$, and this consistent orientation remains when we transform the $xyz$ graph representation of $G_{u,v}$ into an $xyz$ graph representation of $G$.

In the other direction, suppose that $G$ has an $xyz$ graph representation that is consistently oriented. Form the corresponding $xyz$ surface, and transform it into an $xyz$ surface representation of $G_{u,v}$ by cutting out a small disk from the surface surrounding $u$, cutting out another small disk surrounding $v$, and connecting these two disks by a cylinder with six colored faces, as shown in Figure~\ref{fig:contop}. Since $G_{u,v}$ has a valid $xyz$ surface representation, by Theorem~\ref{thm:xyz surface}, it corresponds to a valid $xyz$ graph representation of $G_{u,v}$.
\end{proof}

\begin{figure}[t]
\centering\includegraphics[width=5in]{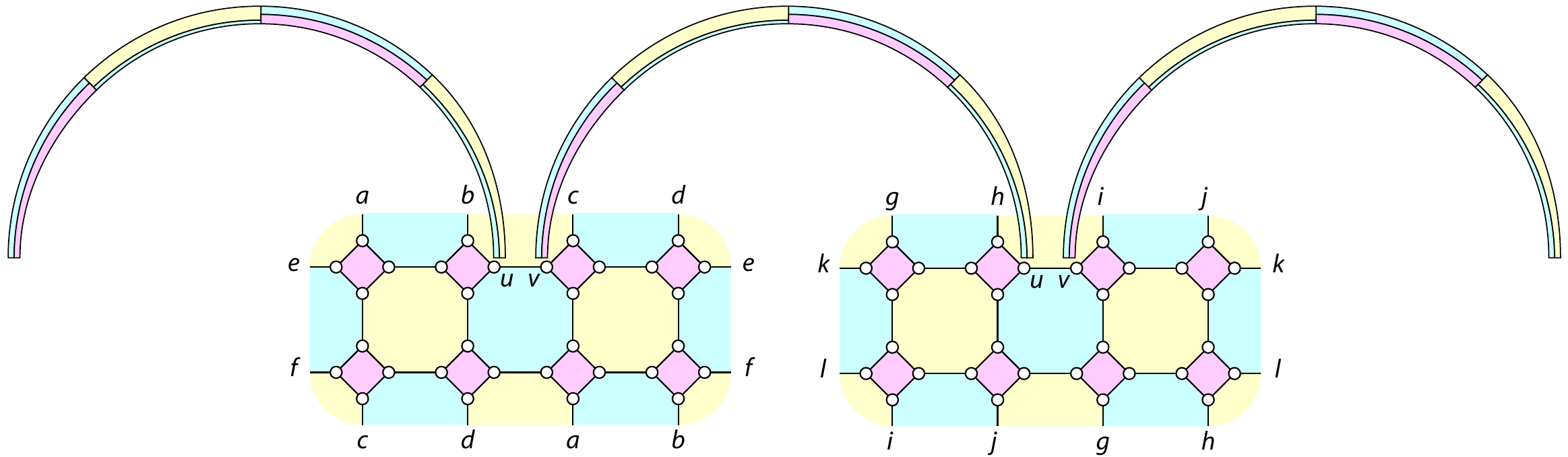}
\caption{Schematic view of edge gadget: three connector gadgets connected sequentially via two flip gadgets.}
\label{fig:edge-gadget}
\end{figure}

\begin{theorem}
It is NP-complete, given an undirected graph $G$, to determine whether $G$ can be represented as an $xyz$ graph.
\end{theorem}

\begin{proof}
Membership in NP follows from Theorem~\ref{thm:test xyz surface}. To prove NP-hardness, we reduce from the standard NP-complete problem of graph 3-coloring.

Thus, given a graph $H$, we wish to construct from it in polynomial time a graph $G$, such that $G$ is an $xyz$ graph if and only if $H$ is 3-colorable. We do so by connecting together subgraphs of various types using the connector gadget described above. In our graph $G$, colors of vertices will be represented by the orientations of edges on connector gadgets. Specifically, if we label the three edges on one side of a connector gadget $A$, $B$, and $C$, then orientations with $A$ parallel to the $x$ axis will correspond to one color, orientations with $A$ parallel to the $y$ axis will correspond to a second color, and orientations with $A$ parallel to the $z$ axis will correspond to a third color. Thus, each color can be represented by two possible orientations, as we do not care which orientation is assigned to $B$ and which to $C$.

To represent a vertex in $H$ with $d$ incident edges, we use any cubic bipartite 3-connected planar graph with at least $d$ vertices, for instance a $d/2$-prism. By Theorem~\ref{thm:planar} each such graph has a representation as an $xyz$ graph that uniquely orients the edges up to permutation of the coordinates. Let $M$ be a matching in this vertex gadget consisting of edges that have the same orientation in the $xyz$ graph representation of the gadget. We will attach connector gadgets to this vertex gadget in such a way that all connector gadgets have their $A$ edges attached to edges of $M$. Therefore, all attached connector gadgets must represent the same color, which may be any of the three available colors.

The next gadget we will use is the 32-vertex graph with two different toroidal $xyz$ surfaces described in the previous section. Let $u$ and $v$ be any two vertices of this graph that are endpoints of an edge $uv$ but do not belong to a 4-cycle in the graph. Then among the $xyz$ graph representations of this gadget, if $u$ has the triple of orientations $(a,b,c)$ (where $a$ is the orientation of edge $uv$), $v$ may have either of the triples of orientations $(a,b,c)$ or $(a,c,b)$. We call this a \emph{flip gadget} as it allows us to flip the orientations of two of the edges without allowing the orientation of the third to change; we will attach connector gadgets at $u$ and $v$.

Our \emph{edge gadget} is formed by the sequential attachment of a connector, flip gadget, connector, flip gadget, and connector (Figure~\ref{fig:edge-gadget}). The right $A$ edge of the first connector and the left $A$ edge of the second connector are connected to the $uv$ edge of the first flip gadget; the right $B$ edge of the second connector is connected to the $uv$ edge of its flip gadget, as is the left $A$ edge of the third connector. It is straightforward to verify that this connection pattern allows the unattached of the connectors to be given any triples of orientations that represent different colors, but not to be given triples of orientations that represent the same color.

Our overall construction of $G$ from $H$ is to create a vertex gadget for each vertex of $H$, an edge gadget for each edge $e$ of $H$, and attach the free ends of the edge gadgets for $e$ to the vertex gadgets for the endpoints of $e$. If $H$ is three-colorable, we may form an $xyz$ surface that corresponds to the coloring within each vertex gadget, and extend it to an $xyz$ surface for all of $G$ by repeatedly applying Lemma~\ref{lem:abstract attachment}; therefore, in this case, $G$ has an $xyz$ surface and an $xyz$ graph representation.
Conversely, if $G$ has an $xyz$ surface, repeated application of Lemma~\ref{lem:abstract attachment} shows that each vertex gadget must be colored as if it were unattached, so we may determine from the $xyz$ surface a color assignment by examining the orientation of the edges within each vertex gadget. Lemma~\ref{lem:abstract attachment}  and the properties of the edge gadget we have constructed force this color assignment to be a valid coloring of $H$.

Thus, we can reduce 3-coloring to testing whether a graph is an $xyz$ graph, in polynomial time. Since testing for being an $xyz$ graph is in NP, and 3-coloring is NP-complete, testing for being an $xyz$ graph is also NP-complete. 
\end{proof}

\section{Conclusions}

We have studied examples, algorithms, topology, and complexity of $xyz$ graph drawing.
Our investigation opens up several avenues for further research:

\begin{itemize}
\item Our construction of an $xyz$ graph from an $xyz$ surface does not specify the ordering of the coordinate values associated with each face, so we may freely permute this ordering, giving drawings with different appearances for a single $xyz$ surface representation. Figure~\ref{fig:permutohedron} illustrates this: the most direct translation of the permutohedron to an $xyz$ drawing has many crossings, while it is possible to find a different permutation of coordinate values that leads to a drawing with no crossings. How difficult is it, given an $xyz$ surface, to find a permutation of coordinate values that minimizes the number of crossings?

\item Our construction of an $xyz$ graph from an $xyz$ surface assigns a unique coordinate value for each polygon in the surface, but in some cases it may be possible to reduce the number of distinct values needed in a drawing by sharing a coordinate between multiple faces. We may define the \emph{volume} of a drawing (assuming integer coordinate values) as the product $(n_{xy}-1)(n_{yz}-1)(n_{xz}-1)$, where $n_{xy}$, $n_{yz}$, and $n_{xz}$ denote the number of distinct coordinate values used by faces of the $xyz$ surface parallel to each of the axis-parallel planes. For instance, again, in Figure~\ref{fig:permutohedron} , there are four faces parallel to two of the planes and six faces parallel to the third, so the volume of a naively formed drawing would be $3\times 3\times 5=45$, but both of the drawings shown only use five distinct coordinate values in the third direction, placing two polygons on the same plane and giving volume $36$. How difficult is it to find the minimum volume $xyz$ graph drawing of a given $xyz$ surface?

\item The rightmost drawing of the permutohedron in Figure~\ref{fig:permutohedron} shows it as the boundary of an orthogonal polyhedron (perhaps suitable for the design of a building). How difficult is it to determine whether such a representation exists for a given bipartite 3-connected cubic planar graph?

\item Our reduction from graph coloring to testing the existence of an $xyz$ coloring produces graphs of high genus: if the graph to be colored has $n$ vertices and $m$ edges, the resulting $xyz$ surface (if it exists) has genus $3m-n+1$. On the other hand, we saw that testing $xyz$ graph representability is easy in the case of genus zero (that is, planar graphs). From these results we might speculate about the existence of a fixed-parameter-tractable algorithm for this problem. Is there an algorithm for finding $xyz$ surface representations of a graph that is polynomial time for any fixed bound on the genus of the surface it finds?

\item For Cayley graphs with one self-inverse and one non-self-inverse generator, the difficulty in finding $xyz$ graph representations is linked to the need to independently orient each cycle formed by the non-self-inverse generator. However, as these graphs are highly symmetric, it seems natural to hope that these cycles may be oriented in a symmetric way that avoids the need for testing all orientations of all cycles. Is there a Cayley graph that may be represented as an $xyz$ graph only by orienting its cycles asymmetrically?

\item Kuperberg's example of the prism shows that our algorithm for testing $xyz$ graph representability using all partitions of the graph into three matchings cannot be improved, unless we avoid some partitions. However, for the prism itself, there are many partitions that can safely be avoided: for an $xyz$ graph representation, we cannot use any partition into three matchings that uses three different orientations in a single quadrilateral. One can also devise similar conditions that restrict the matchings in hexagons and other short cycles of a given graph. Can one take advantage of these forbidden configurations to eliminate some partitions into matchings earlier in the algorithm and reduce its running time?

\item In our discussion of graphs represented by the points with coordinates summing to 0 or 1 in a $k\times k\times k$ grid, we briefly referred to a similar construction of an infinite $xyz$ graph in an infinite three-dimensional grid, isomorphic to the hexagonal tiling of the plane, a graph treated in more detail in another paper~\cite{Epp-GD-08-ids}. To what extent can the correspondence between $xyz$ graphs and $xyz$ surfaces be generalized to infinite graphs? What is the most appropriate way of handling the infinite chains of edges parallel to a single coordinate plane that can arise in the infinite case?

\begin{figure}[t]
\centering\includegraphics[width=2.5in]{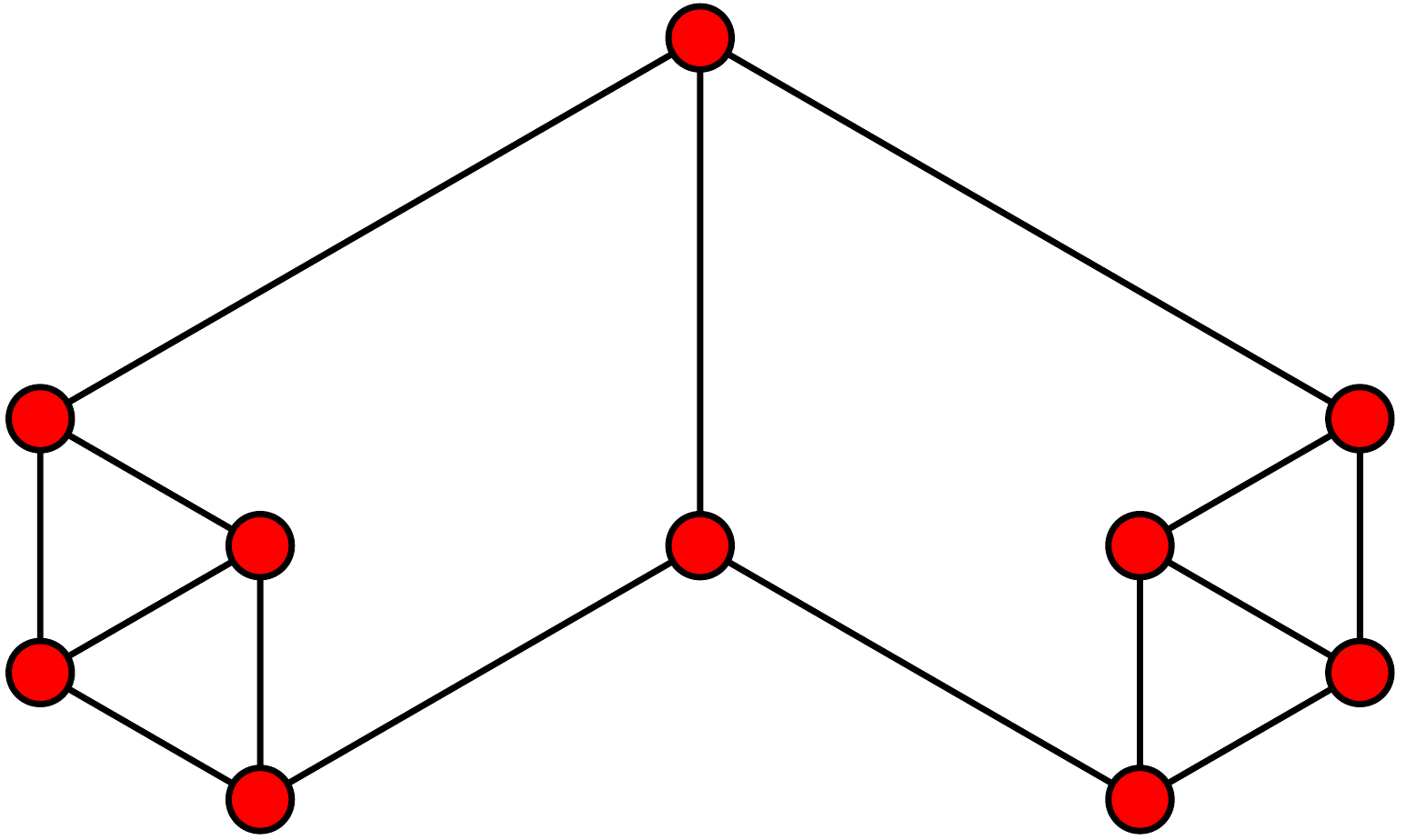}
\caption{A point set such that lines in three parallel families each contain zero or exactly two points, and the cubic graph derived from it. This graph is not an $xyz$ graph, as it contains triangles.}
\label{fig:3-orient}
\end{figure}

\item Any $xyz$ graph can (after a suitable perturbation of its coordinate values) be projected isometrically onto the plane in such a way that all edges are parallel to the sides of an equilateral triangle, and no line through an edge contains any other vertex. More generally, if we have any point set in the plane such that any line parallel to the sides of an equilateral triangle contains either zero or two points, we may define a cubic graph analogously to the three-dimensional definition of $xyz$ graphs.
However, these planar three-orientation graphs are somewhat more general than $xyz$ graphs; for instance Figure~\ref{fig:3-orient} shows a graph that may be drawn in this way that contains triangles and is therefore not an $xyz$ graph. To what extent may our theory be extended to these graphs?
\end{itemize}

\section*{Acknowledgements}

We thank Ed Pegg, Jr., Tomo Pisanski, Frank Ruskey, Tom Tucker, Arthur White, and the anonymous reviewers for Graph Drawing 2008 for helpful comments on an earlier draft of this paper. This work was supported by NSF grant 0830403. Except as otherwise noted, all figures in this paper are by the author; all figures remain the copyright of their creators and are used by permission.

\raggedright
\bibliographystyle{abuser}
\bibliography{xyz}

\begin{thebibliography}{10}

\bibitem{AnnBauRos-90}
F.~Annexstein, M.~Baumslag, and A.~L. Rosenberg.
\newblock {Group action graphs and parallel architectures}.
\newblock {\em SIAM Journal on Computing} 19(3):544{--}569, 1990.

\bibitem{BieSheWhi-JGAA-99}
T.~Biedl, T.~C. Shermer, S.~Whitesides, and S.~K. Wismath.
\newblock {Bounds for orthogonal 3-D graph drawing}.
\newblock {\em J. Graph Algorithms and Applications} 3(4):63{--}79, 1999.

\bibitem{BieThiWoo-Algo-06}
T.~Biedl, T.~Thiele, and D.~R. Wood.
\newblock {Three-dimensional orthogonal graph drawing with optimal volume}.
\newblock {\em Algorithmica} 44(3):233{--}255, 2006.

\bibitem{BonLit-95}
C.~P. Bonnington and C.~H.~C. Little.
\newblock {\em {The Foundations of Topological Graph Theory}}.
\newblock Springer-Verlag, Berlin, Heidelberg, and New York, 1995.

\bibitem{CalMas-TCS-01}
T.~Calamoneri and A.~Massini.
\newblock {Optimal three-dimensional layout of interconnection networks}.
\newblock {\em Theoretical Computer Science} 255(1{--}2):263{--}279, 2001.

\bibitem{ChuRobSey-AM-06}
M.~Chudnovsky, N.~Robertson, P.~Seymour, and R.~Thomas.
\newblock {The strong perfect graph theorem}.
\newblock {\em Annals of Mathematics} 164:51{--}229, 2006.

\bibitem{CloGarJohWis-JGAA-01}
M.~Closson, S.~Gartshore, J.~R. Johansen, and S.~K. Wismath.
\newblock {Fully dynamic 3-dimensional orthogonal graph drawing}.
\newblock {\em J. Graph Algorithms and Applications} 5(2):1{--}34, 2001.

\bibitem{CraWhi-DM-08}
D.~L. Craft and A.~T. White.
\newblock {3-maps}.
\newblock {\em Discrete Mathematics}, 2008.

\bibitem{DujEppSud-CGTA-07}
V.~Dujmovi{\'c}, D.~Eppstein, M.~Suderman, and D.~R. Wood.
\newblock {Drawings of planar graphs with few slopes and segments}.
\newblock {\em Computational Geometry Theory and Applications} 38:194{--}212,
  2007.

\bibitem{EadStiWhi-IPL-96}
P.~Eades, C.~Stirk, and S.~Whitesides.
\newblock {The techniques of Komolgorov and Bardzin for three-dimensional
  orthogonal graph drawings}.
\newblock {\em Information Processing Letters} 60(2):97{--}103, 1996.

\bibitem{EadSymWhi-GD-96}
P.~Eades, A.~Symvonis, and S.~Whitesides.
\newblock {Two algorithms for three dimensional orthogonal graph drawing}.
\newblock {\em Proceedings of the 4th International Symposium on Graph
  Drawing}, vol. 1190, pp.~139{--}154. Springer-Verlag, Lecture Notes in
  Computer Science, 1996.

\bibitem{Epp-SODA-03}
D.~Eppstein.
\newblock {Dynamic generators of topologically embedded graphs}.
\newblock {\em Proc. 14th Symp. Discrete Algorithms}, pp.~599{--}608. ACM and
  SIAM, January 2003, arXiv:cs.DS/0207082.

\bibitem{Epp-EJC-06}
D.~Eppstein.
\newblock {Cubic partial cubes from simplicial arrangements}.
\newblock {\em Electronic J. Combinatorics} 13(1, R79):1{--}14, September 2006,
  arXiv:math.CO/0510263.

\bibitem{Epp-GD-08-ids}
D.~Eppstein.
\newblock {Isometric diamond subgraphs}.
\newblock {\em Proc. 16th Int. Symp. Graph Drawing}, 2008.

\bibitem{EveTar-TCS-76}
S.~Even and R.~E. Tarjan.
\newblock {Computing an $st$-numbering}.
\newblock {\em Theoretical Computer Science} 2(3):339{--}344, 1976.

\bibitem{GaiGup-SJAM-77}
P.~Gaiha and S.~K. Gupta.
\newblock {Adjacent vertices on a permutohedron}.
\newblock {\em SIAM Journal on Applied Mathematics} 32(2):323{--}327, 1977.

\bibitem{GroTuc-87}
J.~L. Gross and T.~W. Tucker.
\newblock {\em {Topological Graph Theory}}.
\newblock John Wiley {\&} Sons, 1987.
\newblock Dover reprint, 2001.

\bibitem{Hea-QJPAM-98}
P.~J. Heawood.
\newblock {On the four-colour map theorem}.
\newblock {\em Quarterly J. Pure Appl. Math.} 29:270{--}285, 1898.

\bibitem{JenTof-95}
T.~R. Jensen and B.~Toft.
\newblock {\em {Graph Coloring Problems}}.
\newblock John Wiley {\&} Sons, 1995.

\bibitem{Koc-GD-08}
M.~Kochol.
\newblock {3-regular non 3-edge-colorable graphs with polyhedral embeddings in
  orientable surfaces}.
\newblock {\em Proc. 16th Int. Symp. Graph Drawing}, 2008.

\bibitem{NelWil-90}
R.~Nelson and R.~J. Wilson.
\newblock {\em {Graph Colourings}}.
\newblock John Wiley {\&} Sons, 1990.

\bibitem{PapTol-JGAA-99}
A.~Papakostas and I.~G. Tollis.
\newblock {Algorithms for incremental orthogonal graph drawing in three
  dimensions}.
\newblock {\em J. Graph Algorithms and Applications} 3(4):81{--}115, 1999.

\bibitem{Pis-NZJM-07}
T.~Pisanski.
\newblock {Yet another look at the Gray Graph}.
\newblock To appear in {\em New Zealand Journal of Mathematics} 36:85{--}92,
  2007.

\bibitem{Pol-MISH-90}
C.~Le~Conte~de Poly-Barbut.
\newblock {Le diagramme du treillis permutoh{\`e}dre est Intersection des
  diagrammes de deux produits directs d'ordres totaux}.
\newblock {\em Math{\'e}matiques, Informatique et Sciences Humaines}
  112:49{--}53, 1990.

\bibitem{PreVui-CACM-81}
F.~P. Preparata and J.~Vuillemin.
\newblock {The cube-connected cycles: a versatile network for parallel
  computation}.
\newblock {\em Communications of the ACM} 24(5):300{--}309, 1981.

\bibitem{Foster}
G.~Royle, M.~Conder, B.~McKay, and P.~Dobscanyi.
\newblock {Cubic symmetric graphs (The Foster Census)}.
\newblock Web page http://people.csse.uwa.edu.au/gordon/remote/foster/, 2001.

\bibitem{Tuc-JCTB-84}
T.~W. Tucker.
\newblock {There is only one group of genus two}.
\newblock {\em Journal of Combinatorial Theory, Series B} 36:269{--}275, 1984.

\bibitem{Woo-GD-98}
D.~R. Wood.
\newblock {An algorithm for three-dimensional orthogonal graph drawing}.
\newblock {\em Proceedings of the 6th International Symposium on Graph
  Drawing}, vol. 1547, pp.~332{--}346. Springer-Verlag, Lecture Notes in
  Computer Science, 1998.

\bibitem{Woo-01}
D.~R. Wood.
\newblock {Bounded degree book embeddings and three-dimensional orthogonal
  graph drawing}.
\newblock {\em Proceedings of the 9th International Symposium on Graph
  Drawing}, vol. 2265, pp.~312{--}327. Springer-Verlag, Lecture Notes in
  Computer Science, 2001.

\bibitem{Woo-TCS-03}
D.~R. Wood.
\newblock {Optimal three-dimensional orthogonal graph drawing in the general
  position model}.
\newblock {\em Theoretical Computer Science} 299(1{--}3):151{--}178, 2003.

\end{thebibliography}
\end{document}